\documentclass[manuscript, table]{acmart}

\usepackage{mathtools}
\DeclarePairedDelimiter{\ceil}{\lceil}{\rceil}
\usepackage{booktabs}
\usepackage{csquotes}
\usepackage{pifont}
\usepackage{soul}
\usepackage{bm}
\usepackage{algorithm}
\usepackage[noend]{algpseudocode}
\usepackage{tikz}

\newcommand{\ra}[1]{\renewcommand{\arraystretch}{#1}}

\definecolor{grayback}{HTML}{D8DCE2}

\usepackage{array}
\AtBeginDocument{%
  \providecommand\BibTeX{{%
    \normalfont B\kern-0.5em{\scshape i\kern-0.25em b}\kern-0.8em\TeX}}}

\copyrightyear{2022}
\acmYear{2022}
\setcopyright{acmcopyright}\acmConference[CHI '22]{CHI Conference on Human Factors in Computing Systems}{April 29-May 5, 2022}{New Orleans, LA, USA}
\acmBooktitle{CHI Conference on Human Factors in Computing Systems (CHI '22), April 29-May 5, 2022, New Orleans, LA, USA}
\acmPrice{15.00}
\acmDOI{10.1145/3491102.3517649}
\acmISBN{978-1-4503-9157-3/22/04}

\begin{document}

\title[]{Supporting Serendipitous Discovery and Balanced Analysis of Online Product Reviews with Interaction-Driven Metrics and Bias-Mitigating Suggestions}






\author{Mahmood Jasim}
\affiliation{\institution{University of Massachusetts Amherst}
\country{USA}}
\email{mjasim@cs.umass.edu}

\author{Christopher Collins}
\affiliation{\institution{Ontario Tech University}
\country{Canada}}
\email{christopher.collins@ontariotechu.ca}

\author{Ali Sarvghad}
\affiliation{\institution{University of Massachusetts Amherst}
\country{USA}}
\email{asarv@cs.umass.edu}

\author{Narges Mahyar}
\affiliation{\institution{University of Massachusetts Amherst}
\country{USA}}
\email{nmahyar@cs.umass.edu}

\renewcommand{\shortauthors}{Jasim, et al.}

\begin{abstract}
In this study, we investigate how supporting serendipitous discovery and analysis of online product reviews can encourage readers to explore reviews more comprehensively prior to making purchase decisions. We propose two interventions --- Exploration Metrics that can help readers understand and track their exploration patterns through visual indicators and a Bias Mitigation Model that intends to maximize knowledge discovery by suggesting sentiment and semantically diverse reviews. We designed, developed, and evaluated a text analytics system called Serendyze, where we integrated these interventions. We asked 100 crowd workers to use Serendyze to make purchase decisions based on product reviews. Our evaluation suggests that exploration metrics enabled readers to efficiently cover more reviews in a balanced way, and suggestions from the bias mitigation model influenced readers to make confident data-driven decisions. We discuss the role of user agency and trust in text-level analysis systems and their applicability in domains beyond review exploration.
\end{abstract}



\begin{teaserfigure}
 \includegraphics[width=\textwidth]{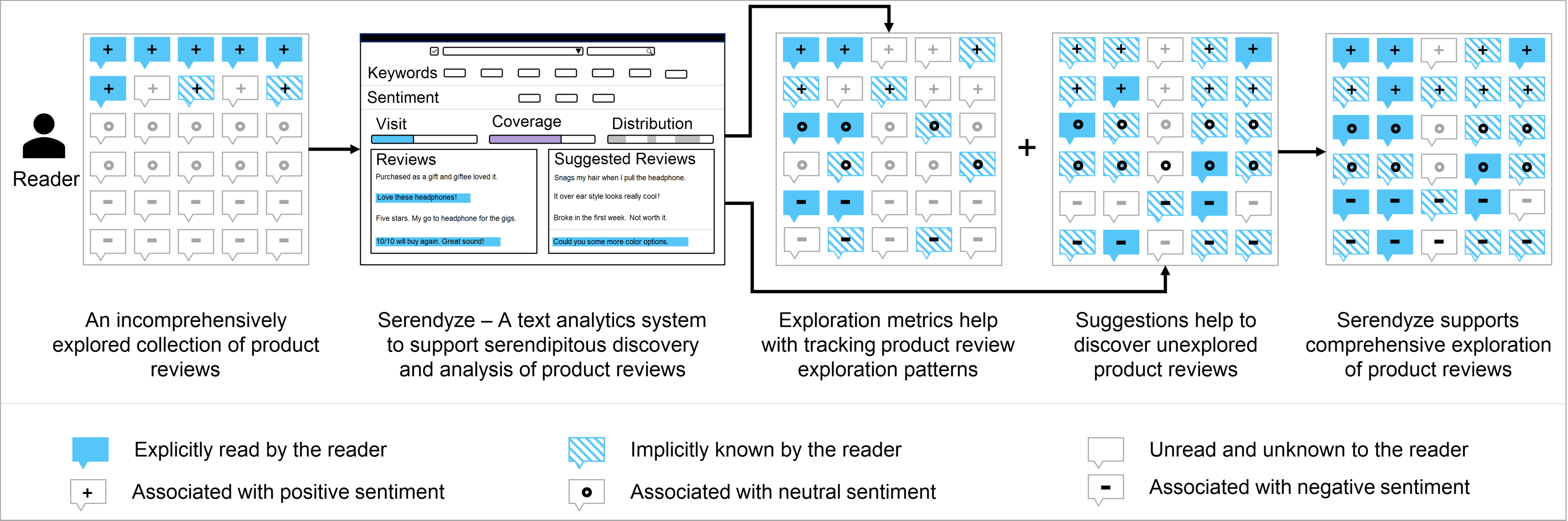}
 \caption{Serendyze is a text analytics system that uses two novel interventions --- exploration metrics and a bias mitigation model --- to enable readers to explore product reviews more comprehensively. The exploration metrics help readers track their data exploration across different facets, such as sentiments. The bias mitigation model suggests reviews that are semantically and sentiment-wise dissimilar to what the readers have been exploring so that they can discover a broader range of reviews. Integrated within an interactive interface, these features can enable readers to gain comprehensive knowledge about the data prior to decision-making.}
\end{teaserfigure}

\maketitle

\section{Introduction}
Customers of online products often depend on product reviews to make data-driven purchase decisions~\cite{von2018influence, hu2014ratings}.
These product reviews --- free-form text comments from previous customers that highlight their opinions and evaluations of online products --- are often considered the most influential factor behind sales and attitudes towards a product~\cite{von2018influence, floyd2014online}. 
While customers might have different strategies to navigate reviews to make their decisions~\cite{chen2016exploring}, those who prefer to comprehensively explore and analyze product reviews often struggle to do so due to the abundance of reviews available~\cite{jones2004information, kwon2015people} and the limited amount of time to accrue insights from them~\cite{park2006information, kwon2015people}. 
As such, these customers are often unable to evaluate all available alternatives in depth~\cite{haubl2000consumer}, which often results in incomplete exploration and understanding of the underlying product reviews~\cite{hu2009overcoming, von2018influence, kwon2015people} prior to making purchase decisions. 

Recent interest in data exploration and discovery~\cite{kunaver2017diversity, lu2015recommender} along with beyond-accuracy metrics~\cite{kaminskas2016diversity} has prompted research into identifying and presenting diverse and serendipitous information to increase people's coverage and understanding of the data. 
Coupled with information visualization research geared towards providing navigational cues to investigate how readers interact with visual artifacts~\cite{willett2007scented, sarvghad2016visualizing, wall2019markov}, serendipitous information\footnote{The term \textit{Serendipity} has been defined in various ways by previous researchers~\cite{thudt2012bohemian, kaminskas2016diversity}. 
In this paper, we define \textit{serendipity} as an \textit{unexpected yet beneficial discovery that adds to the knowledge of the readers about the data they are exploring.}} --- information that is yet unexplored by the readers and may add to their knowledge of the underlying data --- has shown promise in expanding the depth and breadth of data exploration~\cite{sarvghad2016visualizing, wall2019markov, north2011analytic}. However, these existing methods that encourage data exploration by increasing data coverage were not designed for product reviews --- or texts in general --- and their effectiveness on numerical or categorical data may not translate to predominantly text-based product reviews. 

Prior works also suggest that users' innate cognitive biases often influence how they interact with data using analytics systems~\cite{wall2017warning}. 
For instance, people who are oversensitive to consistency~\cite{heuer1999psychology, wall2017warning} tend to interact with data that
supports the broadest encompassing hypothesis, dismissing other data.
When reading product reviews, this bias may influence a reader to read reviews that are predominantly positive or negative~\cite{hu2009overcoming, haubl2000consumer}.
Furthermore, the persistence of impressions based on discredited evidence~\cite{heuer1999psychology, wall2017warning} often results in continuous interaction with data supporting a hypothesis that has been disproved. 
This bias may prompt readers to ignore reviews that highlight issues with their preferred products. 
These biases often manifest when users are overwhelmed with large amounts of data, resulting in them following their preconceptions, anchoring biases, and using biases as filters to explore underlying data~\cite{wall2018four}. This manifestation of innate bias is inadvertently amplified by systems that respond to users' interactions and preferences --- facilitating incomplete, ineffective, and often biased data exploration prior to decision-making~\cite{wall2017warning, wall2019markov}. 

In this work, we investigate interventions that are intended to support serendipitous discovery and analysis to help readers comprehensively read and tease apart valuable insights from free-form texts in a balanced way. Here, we demonstrate how these interventions might work in the context of online product reviews.
To that end, we investigate a two-pronged approach. 
First, we propose three interaction-driven exploration metrics \textbf{Visit} --- a measure of reviews a reader has explicitly interacted with, \textbf{Coverage} --- a measure of reviews covered by a reader implicitly, that are similar and redundant to the reviews they have already visited, and \textbf{Distribution} --- a measure of the relation of reviews the reader has visited from different facets, such as sentiments, to the true distribution of that facet in the dataset.
Second, we propose a \textit{bias mitigation model} to improve knowledge discovery and balance overall review exploration. 
The model tracks how a reader has been visiting reviews and generates suggestions that are semantically and sentiment-wise different from what they have visited already. 

The interaction-driven exploration metrics are designed to act as an awareness mechanism to help readers understand and track their review exploration progress and patterns through visual indicators. 
They highlight which reviews the readers have implicit and explicit knowledge about and the reviews that are left unexplored. 
The bias mitigation model is designed to support serendipitous discovery and offers a complementary view of reviews read.
It is aimed at helping readers to balance their holistic understanding, increase data coverage, and mitigate bias towards specific review sentiments by providing them with suggestions that are different from what they have visited already.

We integrated the exploration metrics and the bias mitigation model with an interactive text analytics system, \textit{Serendyze}. We use Serendyze to investigate the following questions: 
\begin{enumerate}
    \item \textbf{RQ1:} Does supporting serendipitous discovery and analysis help readers to perform in-depth exploration to cover more product reviews? 
    \item \textbf{RQ2:} How do readers’ review exploration behaviors change when they have access to their exploration patterns? 
    \item \textbf{RQ3:} How do suggestions from unexplored reviews impact readers’ online product purchase decisions?
\end{enumerate}

In this study, we used Amazon product reviews as an example dataset. Furthermore, among myriad online products, we selected headphones as the candidate due to their ubiquitous usage~\cite{headphone}. To study how serendipitous discovery and analysis may impact review exploration, knowledge gathering, and decision-making, we conducted a crowd-sourced between-subjects study in which 100 participants used Serendyze to select their most preferred headphones to recommend to someone. 

The findings from our study demonstrate that exploration metrics and bias mitigating suggestions enable readers to make more informed and confident purchase decisions. 
We found that the majority of the participants who used both exploration metrics and suggestions were confident that they visited enough reviews to make an informed decision (16/25) as opposed to the participants who did not use these features (10/25). 
The majority of the participants who used the features were also confident that they had made the right decision (19/25) compared to those who did not (9/25). 

From the collected usage logs, we found that participants who used exploration metrics and bias mitigating suggestions \textit{covered} an average of $234$ reviews before making a purchase decision, with a 12.28 coverage to time-spent ratio. Participants who did not use these features covered an average of only $66$ reviews, with 8.64 coverage to time-spent ratio.
By the term \textit{covered}, we mean the number of reviews the participants have explicit or implicit knowledge about. 
We consider a participant has \textit{explicit} knowledge about a review if they have visited the review by marking it as read and they have \textit{implicit} knowledge about reviews that are semantically similar to the reviews they visited.
The coverage numbers from our study suggest that participants who used exploration metrics and bias mitigating suggestions had a much broader coverage and knowledge of reviews. 

The collected usage logs and responses to the post-study questionnaire also suggest that Serendyze helped the participants (18/25) to gather comprehensive knowledge from reviews by enabling them to visit reviews in a balanced way, without leaning towards specific sentiments (positive, negative, or neutral). We consider a readers' review exploration as ``balanced'' when the sentiments visited by the reader reflect the true distribution of sentiments present in the dataset. 
Furthermore, the participants who used the suggestions discovered reviews of opposing viewpoints that they were unaware of before, which enriched their knowledge about the products and positively impacted their purchase decisions. 

Based on the findings from our study, we highlight our contributions as follows: 
\begin{enumerate}
    \item A novel approach that intends to support serendipitous discovery and analysis using three interaction-driven exploration metrics and a bias mitigation model to help readers more comprehensively explore product reviews prior to making purchase decisions. 
    \item Empirical evidence that demonstrates the utility of an example text analytics system, Serendyze, integrated with functionalities to track review exploration and allow exploration of serendipitous information from product reviews. The system shows reflective metrics to readers about their review exploration patterns and suggests reviews that they might not have considered otherwise to help them accumulate comprehensive knowledge useful for informed decision-making. 
    \item Discussions on how systems designed to support serendipitous discovery and analysis can be useful in combating biased review exploration. We also discuss readers' agency in mixed-initiative systems and the expansion of systems such as Serendyze for data-driven decision-making in domains beyond product reviews.
\end{enumerate}

\section{Related Work}
Prior works have shown that product reviews are among the most important factors that influence sales and attitudes towards a product~\cite{von2018influence, floyd2014online} and the purchase decisions people make online~\cite{hu2014ratings}. 
In 2020, Qualtrics revealed that 93\% of customers mentioned that online product reviews impacted their purchase decisions~\cite{qualtrics}.
This section describes existing tools and techniques for product review analysis and how serendipitous information discovery could support readers with review exploration and understanding. 

\subsection{Visual Analytics Approaches for Online Product Reviews}
Researchers have explored various text analysis techniques such as opinion extraction, sentiment analysis, topic modeling, and trend analysis, and combined them with visualizations to enable exploration and analysis of product reviews~\cite{khan2014mining, alharbi2019sos, allahyari2017text, kucher2015text, dave2003mining}. 
For example, to explore and analyze online product reviews, OpinionBlocks provides an aspect-based summary of product reviews using a block visualization to present an overview of positive and negative reviews~\cite{hu2013opinionblocks}. 
Review Spotlight summarizes user reviews on restaurants using objective-noun pairs organized as tag clouds~\cite{yatani2011review}.

To facilitate the comparison of opinions among different products derived from text mining across various features, Carenini et al. proposed a multimedia interface~\cite{carenini2009multimedia} to aggregate opinions using bar chart visualizations. Opinion Observer is another such system that enables comparison of people's opinions on product features based on opinion mining by summarizing the pros and cons of the product features~\cite{liu2005opinion}. Chen et al. utilized term-variation patterns to identify underlying topics present in product reviews to facilitate understanding conflicting opinions towards online products using a host of visualizations~\cite{chen2006visual}. 

Others have experimented with extracting and presenting affective content from product reviews. For example, Gregory et al. enabled user-directed affective content exploration in product reviews using variations of rose plots~\cite{gregory2006user}. Furthermore, they experimented with thematic clustering based on keyword extraction to enable exploration of product reviews~\cite{gregory2006user}. OpinionSeer enables multilevel exploration of opinion data from hotel reviews with explicit consideration towards uncertainty using augmented radial charts~\cite{wu2010opinionseer}. 

Prior works suggest that many of these methods often focus on providing aggregated statistics and summaries and put less emphasis on comprehensive exploration and knowledge discovery from the actual text~\cite{kucher2015text, alharbi2019sos, khan2014mining}. 
While these methods are useful for making quick purchase decisions based on an overall impression of the product~\cite{haubl2000consumer}, it might be worthwhile to explore alternatives for potential customers who seek to comprehensively explore and analyze product reviews in-depth to identify nuggets of information that might help them to make more confident data-driven purchase decisions.

\subsection{Tools and Techniques to Support Serendipitous Data Discovery and Analysis}
Researchers in recommender systems --- a subclass of information filtering systems --- focused on identifying data items by predicting how a user might rate the item~\cite{ricci2011introduction} across various domains, including and beyond online products. 
Although research in this area has mostly focused on the accurate prediction of user preferences, there has been a recent interest in exploring methods beyond traditional accuracy-based metrics~\cite{kaminskas2016diversity}.
For instance, researchers have explored metrics to diversify data recommendations~\cite{shi2012adaptive}, provide novel recommendations~\cite{zhang2002novelty}, or support serendipitous discovery of data items~\cite{onuma2009tangent}.
Among various beyond-accuracy metrics explored in prior works~\cite{kaminskas2016diversity, herlocker2004evaluating}, serendipity has received significant attention in the last decade. 

The term \textit{serendipity} is often referred to as the process of finding valuable or surprising things that are not looked for~\cite{andel1994anatomy, herlocker2004evaluating}. 
Others have defined serendipity as a combination of surprise and relevance~\cite{herlocker2004evaluating}. 
However, existing methods adhering to such definitions have mainly focused on suggesting relevant data items and rejecting irrelevant ones~\cite{kotkov2016survey, marchionini2006exploratory}, which may lead to neglecting unpopular or marginalized opinions. 
For instance, consider a reader reading product reviews of headphones using a system that suggests reviews to the reader based on relevance. 
If the reader reads reviews that focus on the price, they might receive more, albeit different, suggested reviews about the price. 
They might not be suggested reviews regarding other aspects such as color or sound quality because the system may consider these aspects irrelevant based on what the reader has been reading. 
As a result, the reader might make a purchase decision without learning about other aspects of the headphone that might be important to them. 
In contrast, we consider serendipity to be an \textit{unexpected yet beneficial discovery of information that adds to the readers' knowledge.}
Our goal is to support the serendipitous discovery of unexpected information that could help readers broaden and improve their knowledge acquisition instead of reinforcing their existing preconceptions with relevant data items.

Previous research in data visualization has explored ways to support serendipitous discovery and analysis of data~\cite{alexander2014serendip, dork2012pivotpaths, isaacs2014footprints}. 
For instance, Bohemian Bookshelf provides visualizations for exploring book collections that enable people to discover trends and relations within the collection in a playful manner~\cite{thudt2012bohemian}. 
Another work, Serendip, provides a topic modeling tool with multiple views. It focuses on intermixing different scales of data inquiry and information types by visualizing the relationships between the data items~\cite{alexander2014serendip}. 
Another visualization tool that promotes serendipitous discovery is PivotPaths~\cite{dork2012pivotpaths}. It enables playful and casual exploration of interlinked metadata using visual paths in enticing arrangements to motivate people to explore the information. 
Footprints is another analytics tool that uses multiple interconnected visualizations to help users navigate through news articles~\cite{isaacs2014footprints}. 
Footprints also enables people to tag the data as \textit{Read}, \textit{To Read}, and \textit{Useful} to track exploration progress and data coverage. 

While these tools provide functionalities to support the serendipitous exploration of documents, their effectiveness for exploring relatively large text documents, including academic papers, books, and news articles, may not translate to product reviews, which are relatively shorter and often free-form in nature. 
Furthermore, these tools often enable the exploration and analysis of large text corpora at the summary level. 
For instance, PivotPaths enables serendipitous discovery of relationships between facets such as author name, venue, and keywords, but not the actual text content of academic publications. 
Similarly, Footprints enable serendipitous discovery of topics and other metadata such as dates and sources, but not the text content of documents.
In this work, we investigate how providing serendipitous information at the text level might impact the data exploration and analysis process. 
To do so, we explore how methods that intend to support serendipitous information discovery and analysis in the context of online provide reviews might impact customers' purchase decisions.

\subsection{Approaches to Increase Data Coverage and Avoid Biased Exploration}
Prior work suggests that users of analytics tools are often prone to biases when exploring data~\cite{wall2017warning, ellis2018cognitive}.
While interacting with the data and system artifacts, a user's internal biases and presumptions towards the data can impact the exploration and analysis process~\cite{wall2017warning, wall2019markov, hu2009overcoming}. 
Such biases include \textit{oversensitivity to consistency}~\cite{heuer1999psychology, wall2017warning}, where an analyst tends to interact with data that
supports the broadest encompassing hypothesis, and they dismiss other data.
In the product review domain, this bias may manifest and influence a reader to read reviews that are predominantly positive or predominantly negative based on the aggregation of reviews~\cite{hu2009overcoming, chen2016exploring}. 
Furthermore, biases such as \textit{persistence of impressions based on discredited evidence}~\cite{heuer1999psychology, wall2017warning} influence analysts to continue interacting with data that supports a hypothesis but has been disproved already.  
This bias can influence readers to make biased decisions based on their brand or product preference, even when reviews highlight issues with their preferred products. 
One approach to mitigating such biases could involve exposing the differences between the data a user has explored and the overall characteristics of the complete underlying data, making users aware of their innate biases that might be injected during their data exploration~\cite{wall2017warning, collins2018guidance, hu2009overcoming}. 

Existing systems designed towards combating such biases often provide visual and navigational cues on how the user has been exploring the data and interacting with the system to inform users of potentially biased interactions and exploration~\cite{sarvghad2016visualizing, isaacs2014footprints, wall2019markov}. 
For instance, Sarvghad et al. proposed a visual analytics tool to provide analysis history to highlight the dimension coverage of data dimensions explored by the user~\cite{sarvghad2016visualizing}. 
These data dimensions are comprised of different attributes present in tabular data. 
The tool employed a variation of scented widgets to assist analysts in forming questions based on their past data exploration patterns. 
Wall et al.~\cite{wall2019markov} also experimented and modeled users' potential biased behavior while using scatterplots based on the history of their data exploration patterns. 

While these tools, methods, and experiments shed light on the potential of providing navigation cues to avoid biased exploration and increase data coverage, they are primarily focused on ordinal, categorical, or numerical data.
Furthermore, these tools were not designed to investigate how providing such information may impact readers' knowledge acquisition before making purchase decisions based on product reviews.
As such, the effects of supporting serendipitous discovery and analysis of reviews to help readers explore, cover more information, and gather knowledge prior to decision-making remain largely unexplored.

\section{Serendyze}
\label{sec:system}

Serendyze is designed and developed as an interactive text analytics system that intends to propel readers to explore and analyze product reviews more comprehensively before making purchase decisions.
Here, we describe different components and functionalities integrated with Serendyze along with the exploration metrics and bias mitigating model, which are intended to support serendipitous discovery and analysis of product reviews. 

\subsection{Exploration Metrics}
\begin{figure}
\includegraphics[width=1\textwidth]{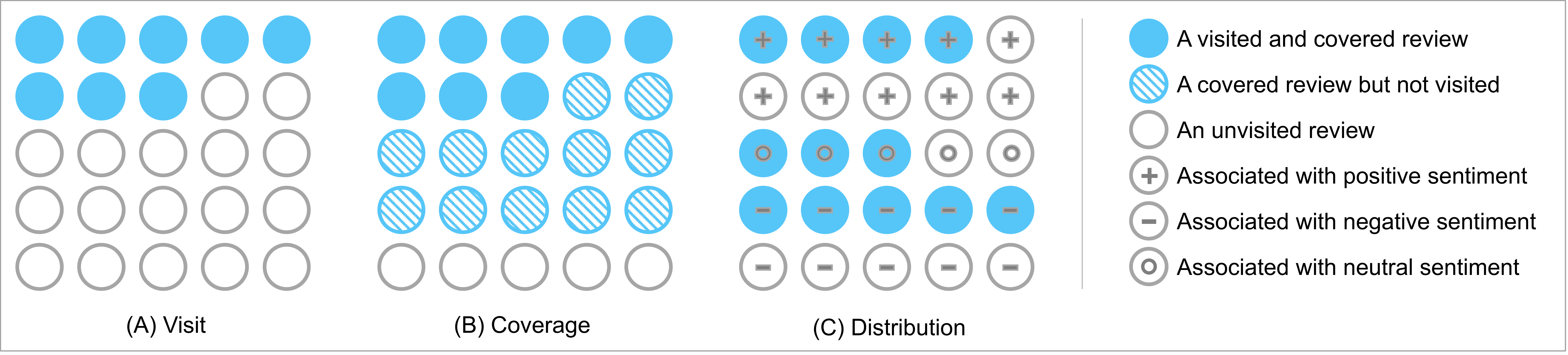}
\caption{Three exploration metrics: A) \textbf{Visit} - a measure of reviews the reader has directly interacted with, B) \textbf{Coverage} - a measure of reviews covered by the reader implicitly through semantic similarity and redundancy, and C) \textbf{Distribution} - a measure of the relation of reviews the reader has visited from different facets, such as sentiments, to the true distribution of that facet. A filled cyan circle represents explicit knowledge: a review the reader has directly interacted with (visited). A striped cyan circle represents implicit knowledge: a review that the reader has not interacted with directly but has \textit{covered} through direct interaction with another semantically similar review. An unfilled gray circle represents a review the reader has not interacted with and has no implicit or explicit knowledge about.}
\label{fig:exploration}
\end{figure}

In this work, we propose three interaction-driven exploration metrics --- \textbf{Visit}, \textbf{Coverage}, and \textbf{Distribution}.
The exploration metrics are designed to enable readers to track their data exploration progress and patterns (see Fig.~\ref{fig:exploration}).

\subsubsection{Visit}
\textbf{Visit} is a measure of reviews a reader has explicitly interacted with. To measure the Visit metric, Serendyze maintains a list of reviews that the reader has marked as read as the \textit{visited list}, $V$. Visit is simply the percentage of reviews marked as read by the reader from the total number of reviews for the product ($N$) using equation~\ref{eq:visit}.  
\begin{equation}
    Visit = \ceil[\Bigg]{\frac{|V|}{N}} \cdot 100
    \label{eq:visit}
    \tag{3.1}
\end{equation}
\subsubsection{Coverage} We define \textbf{Coverage} as a measure of reviews the reader has knowledge of either explicitly or implicitly.
We assume that a reader has \textit{explicit} knowledge about a review if they have visited (read) the review and \textit{implicit} knowledge about a review ($x$) if they have already visited (read) another review ($y$) that is semantically similar to the review ($x$)~\cite{corley2005measuring}.  
For instance, consider two reviews on the same product: \blockquote{\emph{Good Headphones, Great for the price. The headphones work quite well. They don't feel like great headphones but they have held up pretty well and produce good sound.}} and \blockquote{\emph{Great sound, affordable. Great sound for the price and seem like they will last for a while. A good value for the price as well.}}
These reviews are sufficiently semantically similar that they can be considered redundant. 
As such, if a reader visits one of these reviews by marking it as read, we conclude that they have \emph{covered} the other review.
The Coverage metric thus tracks the percentage of reviews the reader has either explicit (visit) or implicit (semantically similar) awareness of. 

To measure Coverage, we first convert each review to a vector representation which embeds semantic information using Doc2Vec~\cite{le2014distributed}.
While Doc2Vec is a generalization of the popular Word2Vec~\cite{mikolov2013efficient} embedding, Doc2Vec's advantage over Word2Vec is its applicability on variable-length documents, making Doc2Vec suitable for embedding product reviews that usually vary in length. 
We decided to use Doc2Vec over other bi-directional language models, such as BERT~\cite{devlin2018bert} and Elmo~\cite{peters2018deep} as it is more interpretable and less computationally expensive~\cite{lau2016empirical} for measuring the semantic similarity among reviews in zero-shot environments.  
However, due to the modular design of Serendyze, Doc2Vec can be replaced with more contemporary transformer-based models for appropriate tasks. 

Serendyze maintains three live lists of Doc2Vec vectors of reviews: visited ($V$), unvisited ($U$), and covered ($C$). 
When a review is visited, pairwise cosine similarity~\cite{gomaa2013survey} between $V$ and $U$ is measured. 
Based on experiments and pilot studies, we use a normalized similarity score of $0.8$ as the threshold to determine if a review is similar enough to be considered redundant and placed in the covered list ($C$).
Any review from the unvisited list with a similarity score of at least $0.8$ to any review from the visited list is added to the covered list ($C_L$).  
Finally, the Coverage value is measured as the percentage of covered reviews --- reviews that a reader has implicit or explicit knowledge about --- from the total number of reviews ($N$) for the product using equation~\ref{eq:coverage}. 
Note that the visited list ($V$) is a subset of the covered list ($C$) as the latter contains all visited reviews with additional redundant reviews.
\begin{equation}
    Coverage = \ceil[\Bigg]{\frac{|C|}{N}} \cdot 100
    \label{eq:coverage}
    \tag{3.2}
\end{equation}
\subsubsection{Distribution}
We define \textbf{Distribution} as a measure of the relation of reviews the reader has visited from different facets, such as sentiments, to the true distribution in the dataset. 
For this study, we considered sentiments (positive, neutral, and negative) as the facet to measure Distribution. 
However, these facets can be customized to include star ratings, sentiments, topics, other metadata, or text mining results. 
In our study, Distribution is a measure of consistency and equilibrium of a reader's review exploration of various sentiments. 
For instance, if a reader focuses heavily on positive reviews while ignoring negative or neutral ones, we consider such exploration patterns  not well-distributed. 
To measure Distribution, Serendyze counts the total number of positive, neutral, and negative reviews for a product. 
During use, Serendyze maintains separate lists of positive, neutral, and negative reviews that a reader has visited. 
As the reader continues to visit reviews, the proportions of visited sentiments are calculated using equation~\ref{eq:distribution}, where $V_X$ is the visited list of sentiment $X$ and $U_X$ is the unvisited list of sentiment $X$. $X$ can be positive, neutral, or negative. 

The Distribution metric is designed to help readers understand how well their visit history reflects the true distribution of sentiments. For example, if a dataset contains vastly more positive reviews than other categories, an unbiased sample of the data would also contain more positive reviews. Because it measures proportions and not the number of reviews, by aiming for the Distribution measure for each sentiment to be equal, the reader could ensure their understanding is reflective of the dataset's true distribution of the sentiments. In this way, the Distribution metric intends to help readers identify if their review exploration is skewed towards a particular category of sentiment and negligent of others. 
\begin{equation}
    Distribution_X = \frac{|V_X|}{|V_X \cup U_X|}
    \label{eq:distribution}
    \tag{3.3}
\end{equation}
While reading reviews, if a reader's Distribution metric for a sentiment exceeds the Distribution metrics of other sentiments by more than 7\%, it is flagged as a tendency to lean towards that sentiment. 
Note that Distribution detects imbalance based on the proportions of reviews visited for each sentiment, not the absolute number. For example, if a reader is focusing on positive reviews to the point where the proportion of positive reviews visited exceeds the proportion of negative and neutral by 7\% or more, we consider the reader's exploration is skewed towards positive reviews. In this way, the measure helps readers stay aware of how their visited reviews reflect the true distribution. 

The threshold of 7\% was determined in pilot testing. We found that a threshold lower than 7\% too aggressively penalized exploration of a certain sentiment. 
In contrast, a higher value allows readers to neglect other sentiments for a longer period.
The strictness of this threshold is fully customizable based on the facets and dataset used. 

\subsection{Bias Mitigation Model}
\label{dismod}
In this work, we propose a heuristic bias mitigation model to extract and present suggestions to readers based on their interactions with reviews.
The model is designed to focus on supporting serendipitous discovery and balanced analysis of reviews by providing suggestions that are intended to encourage readers to visit more reviews and improve their knowledge acquisition about the products. 
To that end, the model suggests unvisited reviews that are semantically and sentiment-wise dissimilar to the reviews the reader has visited already. 
The suggestions are intended to mitigate biased exploration and guide readers to an understanding of the data, which is reflective of the true distributions of the semantic and sentiment diversity in the reviews. 
The complete algorithm to generate the model is presented in Algorithm~\ref{alg:disinterest}.

\begin{algorithm}
\caption{Bias Mitigation Model}
\label{alg:disinterest}
\begin{algorithmic}[1]
\Procedure{Get-Suggestion}{$U, V, S$} \Comment{U, V, S are arrays of unvisited, visited, and visited suggested reviews}
\State $M\gets 1 - \frac{|S_Y|}{|S|}$ \Comment{Calculate score modifiers. $Y\in$ [Dissimilarity, Sentiment], $|M| = 2$}
\State $T\gets\phi$ \Comment{List of objects to store candidate reviews and their scores in tuples}

\For{u in $U$}

\State $d \gets\phi$ \Comment{minimum dissimilarity score}
\State $s \gets\phi$ \Comment{sentiment score}

\State $V' = V + u$ \Comment{Add candidate review to temporary $V$}

\State $P_{pos}\gets |V'_{pos}|/|V_{pos} \cup U_{pos}|$ \Comment{proportion of positive reviews visited}
\State $P_{neut}\gets |V'_{neut}|/|V_{neut} \cup U_{neut}|$ \Comment{proportion of neutral reviews visited}
\State $P_{neg}\gets |V'_{neg}|/|V_{neg} \cup U_{neg}|$ \Comment{proportion of negative reviews visited}

\State $CoV\gets$ Coefficient-of-Variation ($P_{pos}$, $P_{neut}$, $P_{neg}$)  \Comment{Prospective CoV if $u$ is visited}

\State $d \gets 1 - (min(CosineSimilarity(u, V)))$ \Comment{Find max dissimilarity from already visited reviews}
\If {$CoV\:<\:1$} \Comment{Adding $u$ results in distributed reading}
    \State $s \gets 1 - CoV$ \Comment{Give $u$ a higher sentiment score}
     \ElsIf {$CoV\:>\:1$} \Comment{Adding $u$ results in an unbalanced reading}
     \State $s \gets 1 - P_X$ \Comment{$X \in {[pos, neut, neg]}$ associated with $u$}
\EndIf
\If {$CoV\:<\:1 \:\mathbf{\&\&} \: M\:=\:\phi$} \Comment{$M = \phi$, when the reader has not visited any suggestion}
    \State $\mathit{T[u].score}\gets 0.5 * d  + 0.5 * s$ \Comment{Default case} 
     \Else
     \State $\mathit{T[u].score}\gets M[Dissimilarity] * d + M[Sentiment] * s$
\EndIf
    \State $\mathit{T[u].review}\gets u$
     \If {$s > d$} \Comment{Store the dominating component for choosing the suggestion}
     \State $\mathit{T[u].component}\gets Sentiment$
     \Else
     \State$\mathit{T[u].component}\gets Dissimilarity$
     \EndIf
\EndFor
\State Sort($\mathit{T}$) by $\mathit{T.score}$
\State $\mathit{Suggestions}\gets \mathit{T[0:5)}$ \Comment{The first five elements of candidate review list}
\State \Return $\mathit{Suggestions}$ 
\EndProcedure
\end{algorithmic}
\end{algorithm}

There are two major components of the model: (1) The \textbf{dissimilarity measure} that calculates how dissimilar the suggestion is from the reviews that the reader has already visited and (2) The \textbf{sentiment measure} that calculates if the reader is focusing too much on a specific sentiment and neglecting others. The algorithm is called to generate bias mitigating suggestions for every review visited and marked as read by the reader using Serendyze. 
Serendyze maintains several lists, including lists of Doc2Vec vectors of visited ($V$) and unvisited ($U$) reviews, and a list of suggestions the reader has visited ($S$). The list of visited suggestions also contains flags about the primary reason a suggestion was made (to maximize dissimilarity or unbias sentiment). 

For each prospective suggestion $u$, the projected distribution of sentiments is calculated (lines 7--10). Serendyze calculates the coefficient of variation ($CoV$, line 11), a measure of relative variability measured by the ratio of standard deviation to the mean of the visited review proportions of different sentiments.  
A coefficient of variation of less than 1 indicates that the reader is exploring reviews of different sentiments in a distributed fashion. Higher values indicate a greater degree of variability and unbalanced exploration. 
Then, Serendyze calculates pairwise cosine similarity measurement from ($u$) to every review in the visited list ($V$) to generate a maximum dissimilarity score.

When suggesting $u$ would not result in a high $CoV$, the sentiment score $s$ for $u$ is assigned as $1 - CoV$ (lines 13--14). This results in a relatively high score of $s$, which is appropriate as suggestions that do not introduce sentiment distribution biases are preferred. When suggesting $u$ would unbalance the sentiment distribution $CoV > 1$, the sentiment score $s$ is inversely related to the proportion of $u$'s sentiment already visited. As a result, unvisited reviews with sentiments that have not been visited (lower proportion value) will now be scored higher. 

For example, if a reader has been exploring too many positive reviews, they will gradually start to receive negative and neutral reviews as suggestions.
This will increase the chances of an unvisited review with a potentially neglected sentiment to be ranked higher by the model, increasing the reader's chance of receiving diverse suggestions. The final score is a weighted combination of $s$ and $d$. The default weighting is equal (line 18), and the adjustment of weighting factors is discussed below. The top 5 scoring suggestions are returned.

To balance between the two major components of the model, so that one component does not dominate the other while ranking unvisited reviews as suggestion candidates, Serendyze calculates two score modifiers ($M$), where $M[Dissimilarity]$ is the dissimilarity modifier and $M[Sentiment]$ is the sentiment modifier. 
When a reader visits suggestions, the modifiers track the proportion of suggestions that were primarily made for each component ($\frac{|S_Y|}{|S|}$, where $Y \in [Dissimilarity,  Sentiment]$) (line 2). The primary component guiding a suggestion is set on lines 22--25. 

Thus, once some suggestions have been visited, the default scoring formula is replaced with modifier values (line 20). 
With these modifiers, the unvisited reviews are scored in a way that ensures that one component will not dominate the scores. 
For example, if a reader is visiting suggestions whose scores are dominated by dissimilarity, the sentiment modifier ($M[Sentiment]$) will gradually increase in value and start to dominate the score. 
As a result, the reader will receive suggestions geared towards different sentiments from what they have been visiting instead of the semantic dissimilarity of visited reviews. This extension is critical for the readers to receive diverse suggestions that support serendipitous review discovery and develop an unbiased understanding of reviews. 

\subsection{User Scenario}
We present an example scenario to motivate the design and integration of exploration metrics and bias mitigation model with Serendyze. 
Consider Naomi, who is planning to purchase headphones for her brother as a present. 
She wants to find the best option within her limited budget. So, she prefers to explore headphones online with many available options and product reviews to evaluate their values.  
However, from her previous experiences of purchasing products online, she lacks confidence in gathering enough knowledge about different headphones to make the right decision.  

Naomi decides to use Serendyze to explore headphone reviews.
She starts by selecting a headphone. 
Then she reads several reviews and marks them as read. 
While looking through the suggestions, she finds one that talks about the value of the headphone given the price point. 
She hovers over the Coverage bar and finds out from the scented widgets embedded within the keywords that she has not visited any reviews regarding the headphone price. 
She uses the appropriate keyword to filter reviews that mention price. 
At some point during the exploration, she realizes by looking at the Distribution bar that she has been mostly visiting positive reviews. 
She filters the reviews by Negative and finds reviews that show the deficiencies of the headphone, balancing out her overall impression of the headphone. 
Since Serendyze keeps a record of her review exploration, she keeps switching between different headphones and learns more about them without the risk of losing her exploration progress.
She gradually narrows down to a headphone best suited for her needs. 
She hovers over the metrics bars and sees that she has covered aspects important to her, and she has also visited a balanced distribution of positive, neutral, and negative reviews. 
She proceeds to purchase the headphones with confidence that she is informed enough about different headphones to make the best decision. 

\begin{figure}
\includegraphics[width=1\textwidth]{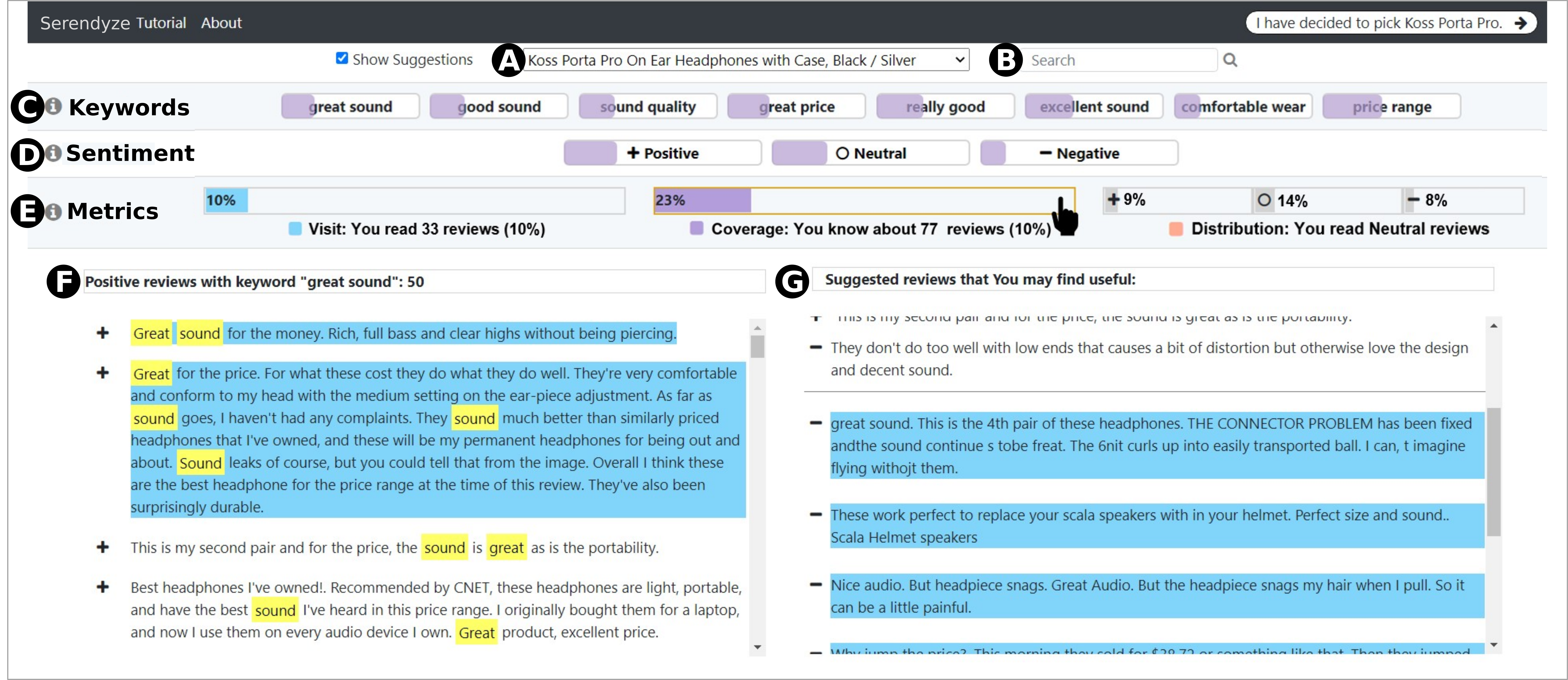}
\caption{Different components in the Serendyze interface: A) a dropdown option for selecting a product, B) a search bar to search for any word present in the reviews, C) a set of filters corresponding to representative keywords, D) filters for positive, neutral, and negative reviews, E) the exploration metrics - Visit, Coverage, and Distribution, F) all product reviews and G) suggested reviews generated by the bias mitigation model that the readers may find interesting.}
\label{fig:Serendyze_interface}
\end{figure}

\subsection{System Description}
Serendyze is an interactive text analytics system designed and developed to help readers explore, analyze, and gather knowledge from product reviews.
Serendyze is designed to support customers who approach reviews in an exploratory manner, as opposed to those who make decisions based on strong personal preferences, such as brand affinities. 
It is intended to help readers who have not decided to purchase a product and want to comprehensively explore options before doing so. 

We compartmentalized the Serendyze interface into several components, including a dropdown option for selecting a product (Fig.~\ref{fig:Serendyze_interface}(A)), a search bar to search for any word present in reviews (Fig.~\ref{fig:Serendyze_interface}(B)), a set of filters corresponding to the most frequently occurring keyword pairs (Fig.~\ref{fig:Serendyze_interface}(C)), filters for positive, neutral, and negative reviews (Fig.~\ref{fig:Serendyze_interface}(D)), the exploration metrics including Visit, Coverage, and Distribution (Fig.~\ref{fig:Serendyze_interface}(E)), and finally, two sets of reviews --- all product reviews (Fig.~\ref{fig:Serendyze_interface}(F)), and suggestions generated by the bias mitigation model (Fig.~\ref{fig:Serendyze_interface}(G)). 
In this section, we describe the functionalities of these components. 

\subsubsection{Keywords and Search}
Serendyze extracts keywords from reviews by identifying all word pairs that co-occur at the document level, where a document is one complete review. 
For visual clarity, we used the top-8 most frequent word pairs as representative keywords for each product. 
These keywords can be used as filters to explore relevant reviews (Fig.~\ref{fig:Serendyze_interface}(C)). 
Serendyze extracts relevant reviews by performing an approximate string search~\cite{baeza1996faster} to identify reviews that contain one or both words from the keyword pairs~\cite{chang2009reading}. 
After filtering the reviews, it highlights all occurrences of the words present in the selected keyword pair (Fig.~\ref{fig:Serendyze_interface}(F)). 

The Search functionality is implemented as an extension of the keyword filters. 
The readers can use the search bar (Fig.~\ref{fig:Serendyze_interface}(B)) to search for any word that might be present in the reviews for the selected product. Upon a successful hit, Serendyze filters the reviews based on the search query and highlights the search word in the reviews. 

Serendyze is designed to be modular and customizable with the option to be outfitted with contemporary topic modeling and keyword extraction methods~\cite{jelodar2019latent}. 
However, due to their probabilistic nature, potential uncertainties  in such systems might pose a threat as a confounding factor. 
As such, we decided to follow a deterministic and explainable method to extract keywords. 

\subsubsection{Sentiments}
\label{sents}
In Serendyze, each review is considered as an individual document, and the reviews were categorized using the associated star rating at the document level. 
We categorized reviews that gave the product 1-star or 2-star rating as negative ($\boldsymbol{-}$), 3-star rating as neutral ($\boldsymbol{\circ}$), and 4-star and 5-star rating as positive ($\boldsymbol{+}$).
Prior works suggest a close interplay between star ratings and product reviews~\cite{tsang2009star} and show that they are highly correlated~\cite{ganu2009beyond}. 
While Serendyze could be outfitted with an off-the-shelf, state-of-the-art, or novel sentiment analysis method~\cite{yadav2020sentiment}, to avoid algorithmic misclassification and maintain transparency, we refrained from using automated sentiment analysis and used star-ratings as user-defined deterministic indicators of valence towards the products. 
However, we did not use the star rating directly as facets, nor incorporated them into the interface directly, because previous studies have also shown that when presented visually, star ratings have an undue cognitive impact compared to sentiments~\cite{schreck2019online}. 
For instance, a review with two visible stars might be perceived more negatively than a positive review with four visible stars~\cite{schreck2019online}. 
We do not claim that star ratings are wrong or unreliable. However, they are not appropriate to be presented visually in our study, as we intended to avoid adding visualizations that might impose additional cognitive impact and distract users from exploration metrics visualization.  
In addition to keywords, these positive, negative, and neutral sentiments associated with reviews can also be used as filters (Fig.~\ref{fig:Serendyze_interface}(D)).


\begin{figure}
\includegraphics[width=1\textwidth]{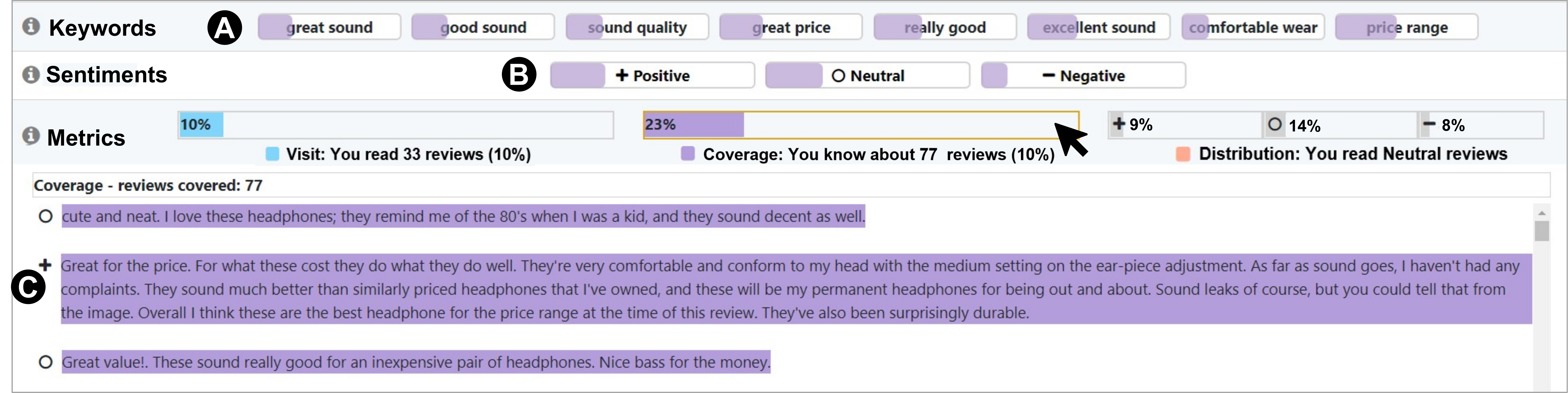}
\caption{Hovering over the Visit or Coverage metric bars reveals data exploration scented widgets embedded in the keyword (A) and sentiment filters (B). Here, the reader hovered over the Coverage bar, and the scented widgets show the keyword pair ``comfortable wear'' and the negative reviews are underexplored compared to other keyword pairs and sentiments. Clicking on the bar filters shows reviews (C) relevant to the exploration metric selected.}
\label{fig:Serendyze_interaction_hover}
\end{figure}

\subsubsection{Exploration Metrics}
In Serendyze, we present three interaction-driven exploration metrics --- \textbf{Visit}, \textbf{Coverage}, and \textbf{Distribution} --- using a set of bar charts.
Readers can use these bar charts to access their data exploration patterns. 
We used horizontal bar charts to visualize Visit and Coverage metrics as they represent percentage values for data visits and data coverage.
We represent Distribution using a set of bar charts that depict the proportion of available positive, neutral, and negative reviews visited by the reader (Fig.~\ref{fig:Serendyze_interface}(E)). 
Each exploration metrics bar is annotated with an explanation of the reader's exploration patterns.
For example, in Fig.~\ref{fig:Serendyze_interaction_hover}, the text below the Visit bar suggests that the reader has explored 33 reviews which is 10\% of the total reviews for this particular product and the text below Distribution suggests that while reading 33 reviews, the reader has been focusing mostly on Neutral reviews.  

The Visit and Coverage bars can be interacted with in two ways. 
First, hovering over these bars transforms the keywords and sentiment filters into scented widgets~\cite{willett2007scented}, providing visual cues of exploration metrics for each keyword pair and sentiment category. 
For example, in Fig.~\ref{fig:Serendyze_interaction_hover} when the reader hovered over the Coverage bar, the keyword pairs (A) and sentiments (B) filled up corresponding to the reader's exploration progress at that time. 
Serendyze follows the visual information seeking mantra~\cite{Shneiderman96theeyes} to trigger the scented widgets on demand to reduce interface clutter and avoid cognitive overload when delivering visual information. 
The second interaction allows readers to drill down and read relevant reviews in detail by clicking on the metrics bars. 
For example, clicking on the Coverage bar allows readers to filter and see the reviews covered  (Fig.~\ref{fig:Serendyze_interaction_hover}(C)).            

\subsubsection{Product reviews.}
Serendyze provides two sets of reviews --- all reviews from the selected product (Fig.~\ref{fig:Serendyze_interface}(F)), and a set of 5 suggestions  (Fig.~\ref{fig:Serendyze_interface}(G)) generated by the bias mitigation model. 
The model is intended to promote serendipitous discovery and analysis of product reviews by providing readers with suggestions that introduce them to features, attributes, or other knowledge related to the selected product that they have not considered or experienced at a rate representative of the true data.

In this work, we used a heuristic bias mitigation model as proposed in~\ref{dismod} to suit our study objectives. 
However, we developed Serendyze as a modular and customizable platform where the bias mitigation model could be replaced with another model that could be used to generate suggestions suitable for other study tasks and domains. 
For example, the heuristic bias mitigation model used in this study that focuses on semantic and sentiment-wise dissimilarity could be replaced with a neural model to suggest similar, popular, or relevant reviews. 

The exploration metrics and bias mitigation model rely on the reader to mark the reviews they have visited already. 
We enabled two ways to mark a review as read. 
The readers can click on any review or hover over a review to mark it as read. 
Previous research on user interaction with interface artifacts suggests that mouse movement is correlated with eye-tracking~\cite{rodden2007exploring, demvsar2017quantifying, sedlar2007tracking}. 
They also suggest that readers are often prone to hovering instead of clicking with interface artifacts~\cite{ha2020expectation}, probing us to include such an alternative. 
The amount of time needed to hover over a review to mark it as read is dynamic and depends on the length of the review. 
In this work, we used a dynamic range from 1 sec to 5 sec to register the hover time to mark a review as read based on the average reading speed of adults~\cite{rayner2010eye} and the length of each review. 
When a reader marks a review as read, the bias mitigation model is called, and Serendyze renders an updated set of suggestions.
Based on the feedback from the pilot study, we retained the suggested reviews that are marked as read below the new suggestions in a chronologically descending order to enable users to keep track of their work (Fig.~\ref{fig:Serendyze_interface}(G)). 
Serendyze saves a readers' review exploration by session.
As a result, switching between products does not remove the reviews marked as read.

\subsection{Implementation Details}
We developed Serendyze as a web application with an HTML, CSS, and JavaScript front-end and a Python backend. 
The Doc2Vec embedding, cosine similarity, and other natural language processing functionalities, including identifying representative keyword pairs, are calculated using the gensim library~\cite{rehurek_lrec}.
The Python scripts were hosted in a freely available server~\cite{pythonanywhere}. 
Upon interaction with the reviews, the front-end fires a request with the list of visited and unvisited reviews, and the server returns the coverage, distribution, and suggestions. 

The system's scalability is dependent on the number of reviews per product. 
We stress-tested the system with over 10000 reviews across 10 products, each containing over 1000 reviews. 
Measured over 100 attempts, it takes Serendyze an average of $3.11 \pm 0.85$ seconds to return suggestions for a product with 1000 product reviews. 
We performed the tests on a laptop with an Intel Core i5 7th generation processor (7300HQ) and 8 gigabytes of RAM, running on localhost. The source code is publicly available for viewing\footnote{\protect{https://osf.io/jmqx2/?view\_only=144115224a204dea8e2104cb829b9606}}. 

\subsection{Pilot Study}
\label{pilot}
Before deploying Serendyze in the real world to study how people use the system to explore online product reviews, we performed a pilot study simulating the same experience with 12 participants (Pi-1 to Pi-12). 
We recruited participants (8 males and 4 females, $28 \pm 4$ years of age on average) using word of mouth and email across different countries.
The goal of the two-week-long pilot study was to simulate and assess the system workflow, identify potential interface issues, and whether participants could use the functions provided in Serendyze to explore the data comprehensively. 

This pilot study helped us to better realize and solidify operational procedures to perform real-world deployment of Serendyze. 
Based on the feedback from the pilot study, we modified the system interface and tuned the threshold values for similarity and distribution. 
We modified the interface by revising the Distribution visualization and used three distinct bars to represent the true distribution of positive, neutral, and negative reviews instead of one aggregate value. 
We also added functionalities to display the already visited suggestions in chronologically descending order below the newly generated suggestions. 
Finally, we fixed several interaction issues, including adding a loading symbol to provide visual feedback that the bias mitigation model is generating new suggestions. We also adjusted the hover time needed to mark a review as read and made other small improvements. 

\section{Evaluation}
To evaluate the viability of supporting serendipitous discovery and analysis of product reviews using exploration metrics and bias mitigating suggestions, we performed a user study with 100 crowd workers. 
The study was approved by the institutional review board.  
In this section, we explain the study conditions, participants, procedure, and findings.

\begin{figure}
\includegraphics[width=1\textwidth]{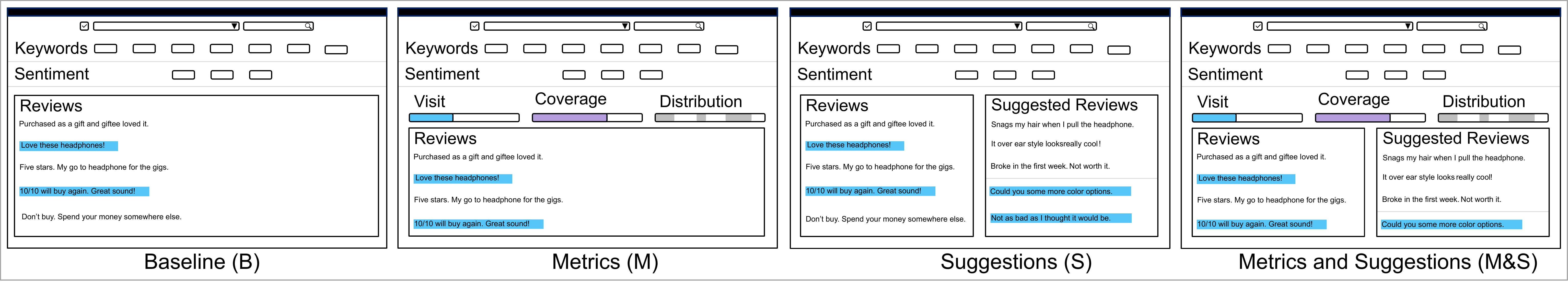}
\caption{This figure depicts all Serendyze components and which features were available with the four conditions (B, M, S, and M\&S). The dropdown for product selection, search options, keywords and sentiment filters, and reviews were available for all four conditions (B, M, S, and M\&S). The exploration metrics (Visit, Coverage, and Distribution) were only available for conditions M and M\&S. The suggestions generated by the bias mitigation model were only available for conditions S and M\&S.}
\label{fig:conditions}
\end{figure}

\subsection{Conditions}
The study was between subjects with four conditions, as presented in Fig.~\ref{fig:conditions}.
Condition B is the \textit{baseline}, condition M is the Serendyze version with exploration \textit{metrics} only, condition S is the Serendyze version with the \textit{suggestions} only, and condition M\&S is the Serendyze version with both \textit{metrics and suggestions}.
Each of these conditions has a set of basic components in common --- the option to select products, the representative keywords, the positive, neutral, and negative sentiment categories, and the reviews.
All conditions enabled users to filter reviews based on keywords and/or sentiments and mark reviews as read by clicking or hovering on them. 
 
We designed conditions M and S to remove confounding factors by evaluating features independently. 
Condition M\&S is the culmination of the Serendyze system with all functionalities. 

\subsection{Participants}
We recruited crowd-worker participants through Amazon Mechanical Turk~\cite{crump2013evaluating}, a popular crowdsourcing platform used to conduct studies requiring human intelligence. 
All of our participants were from North America and were \textit{Amazon Master Workers} who received the qualification for consistent demonstration of a high degree of success in performing a wide range of tasks across many requests. 
25 Master Workers were assigned to each condition. 
Each participant was compensated with USD \$15. 

We asked participants to fill out a pre-study questionnaire to help us understand their online shopping practices and preferences. The response to the questionnaire suggested diverse shopping practices across our participants. 
The overwhelming majority of our participants were familiar with purchasing products online, as 99/100 participants mentioned they purchased at least 1 online product weekly.
Out of 100 participants, only one had never purchased a product online, 37 participants purchased 1--5 products per week, while 31 participants purchased 6--10 products, and another 31 participants purchased more than 10 products. Furthermore, 22 participants spent less than 10 minutes, 29 participants spent 10--20 minutes, 22 participants spent 20--30 minutes, and a final 27 participants spent more than 30 minutes reading product reviews before making purchase decisions. 8 participants were a little dependent on product reviews, while 43 were moderately dependent and 42 were significantly dependent. Another 7 participants were completely dependent on product reviews. 

\subsection{Dataset}
In this study, we used a subset of publicly released Amazon product reviews~\cite{he2016ups} as an example corpus to evaluate Serendyze. 
Among numerous products, we selected headphones as the candidate due to their ubiquitous usage~\cite{headphone}.
Among thousands of headphones in the dataset, we selected three random headphones with over 5000 reviews each, with an average star rating between $4.5$ and $4.6$. 
We chose these conditions to select popular headphones that are not obviously superior or inferior to each other. 
These three headphones were Koss Porta Pro, Sony MDRV6 Studio, and Sennheiser HD280PRO. 
We removed all reviews that contained HTML content or languages other than English. As mentioned in Section~\ref{sents}, we assigned these reviews to positive, negative, and neutral sentiments based on the associated star ratings. 
To keep the number of reviews reasonable for the study participants to read and make decisions, we randomly sampled 120 reviews each for positive, neutral, and negative sentiments for each headphone, aiming for approximately 1000 reviews in total. 
Based on previous work~\cite{woolf2014statistical} and Amazon's product review guidelines~\cite{amguide}, we then removed reviews that are less than 10 words long and more than 100 words long to maintain the length of the reviews at a reasonable level --- suitable for the participants to read and make timely decisions. 
Finally, we ended up with 880 reviews with 338 reviews for Koss Porta, 277 reviews for Sony, and 265 reviews for Sennheiser. 
There were 340 positive, 277 neutral, and 263 negative reviews in total across all three headphones. 
Our sampling did not follow the actual distribution of sentiments present based on star ratings since the headphones chosen were rated mostly positively, and that would result in a dataset with too few neutral and negative reviews. This would make it difficult to reasonably study participants' review exploration across different sentiments. 
Rather, the dataset was constructed to match the study design so that participants could not immediately distinguish among three headphones based on sentiment distribution, and they had to rely on reading reviews to make their decision. 
We used the same dataset with all four conditions. 

\subsection{Procedure}
We asked participants to explore reviews of each of the three headphones using assigned versions of Serendyze and make a decision to refer one of the headphones to someone they know.
We asked them to recommend one of the headphones to others instead of buying for themselves in an attempt to motivate participants to learn about these headphones beyond personal preferences. 
We randomized the procedure of assigning conditions to participants by providing a single link to all crowd workers who participated in the study. 
This link would then redirect the participant randomly to one of the four conditions. 
We also kept a record of studies performed with each condition, and when a condition reached 25 studies, we randomized the remaining redirections to the conditions still not exhausted. 

Each study procedure began with participants' agreement to sign the consent form. 
After signing the consent form, participants were asked to answer a pre-study questionnaire that asked questions about their prior online product review exploration and purchase experiences, including the time spent, the number of reviews read, and products purchased. 
We also asked their favorite headphone brands or feature preferences to see if they influenced their decisions. 

After the pre-study questionnaire, we directed the participants to the tutorial section, featuring a recorded video tutorial explaining the procedures and functionalities of the Serendyze condition assigned to them. 
These videos lasted up to 3 minutes, depending on the condition. Participants could rewind the video but could not skip forward. 
At the end of the video tutorial, the participants proceeded to the study task.
An extended tutorial with annotated figures was also provided to participants and was accessible anytime from the navigation bar. 
In both tutorials, we only presented the features and functionalities for the condition's components. 
We did not disclose the goal of our study or demonstrate any pre-defined exploration patterns to avoid biasing participants' review exploration. 

We instructed the participants to thoroughly read the reviews for all three headphones during the study and decide on a product to refer to others. 
They were also instructed to spend at least two minutes on each headphone.
During these instructions, we did not inform the participants about the goal or hypotheses of the study and did not provide them with any pre-defined exploration patterns or hints. In contrast, the participants were instructed to explore the reviews in any way they wanted, using the features provided in their study condition to make recommendation decisions. 
After deciding, we asked them to finalize their decisions and proceed to the post-study questionnaire. 

In the post-study, we asked participants open-ended questions to learn about their experiences using Serendyze to explore reviews before decision-making. 
In the post-study questionnaire, we added attention checks to identify whether the participants' answers matched their activity during the study. 
We also asked them questions about their usage of exploration metrics and suggestions, their ease of use, how useful they found exploration metrics and suggestions, and how they utilized them while exploring reviews. 
Furthermore, we asked them what they liked and disliked and what issues they faced while working with Serendyze. 
For both the pre-and post-study questionnaires, we asked the participants to answer all questions and ensured that they passed the attention checks before compensating them for their participation. 
The attention checks included questions to verify if the participants could recall information about the headphones. 
All participants passed the attention checks and were compensated.

\subsection{Data Collection and Analysis}
We collected usage logs containing the participants' timestamped interactions with all components of Serendyze and stored them for later analysis. 
We collected the participants' responses to the pre-and post-study questionnaires, the time they spent answering the questions, and the time they spent on the study. 
We embedded the questionnaires with the study platform so that the participants did not have to traverse multiple websites to participate in the study. 

We analyzed the collected data both quantitatively and qualitatively.
We used parametric and non-parametric inferential statistics to analyze the quantitative data.
We analyzed the qualitative data collected from the responses to pre-and post-study questionnaires using an open-coding method~\cite{burnard1991method}. 
Two coders separately and independently coded the questionnaire data collected from condition M\&S as it contains all interventions and potentially most variable data. 
To do so, the coders used spreadsheet applications (Google Sheets and Microsoft Excel) to perform iterative coding in a structured manner. They treated each pre-and post-study questionnaire and their responses individually and performed multiple passes on each response and assigned codes reflective of these responses.
Once all responses were coded by both coders separately, they 
discussed and reached an agreement to consolidate their codes into a representative set of codes. 
The inter-coder reliability measured using Krippendorff's alpha~\cite{krippendorff2011computing} was $0.86$. 
Based on these codes, one coder coded the remaining data collected from conditions B, M, and S, and the other coder verified the codes. 
The data collected from pre-and post-study and the codes for qualitative analysis are provided as supplementary materials. 

\begin{figure}
\includegraphics[width=1\textwidth]{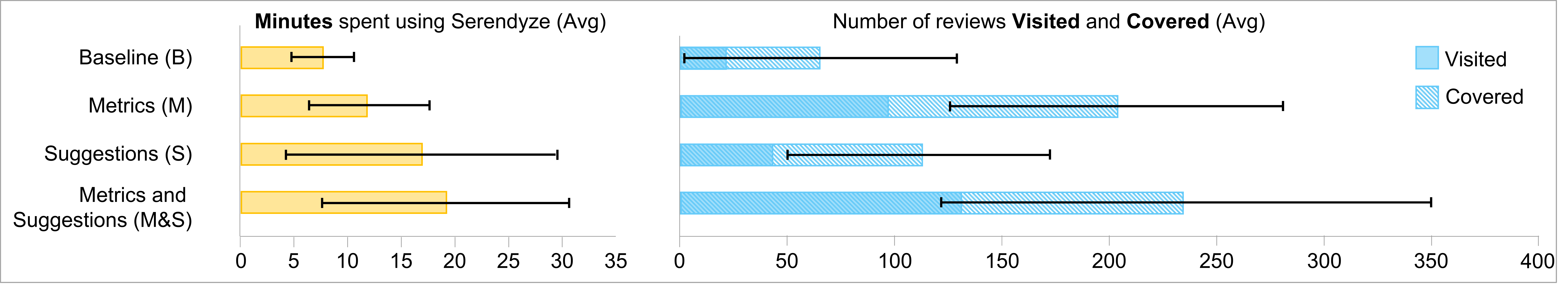}
\caption{Statistics on how long participants used Serendyze and their knowledge about the data. Participants who used condition M\&S spent the most time ($19.05 \pm 11.46$ minutes on average) reading reviews with Serendyze. Participants who used condition M\&S visited the most reviews ($130 \pm 88.87$ reviews on average) and had explicit knowledge about these reviews. Participants who used condition M\&S covered the most reviews ($234 \pm 113.9$ reviews on average) and have implicit knowledge about these reviews.}
\label{fig:Serendyze_usage}
\end{figure}

\subsection{Findings}
\label{sec:results}

Our research questions investigate whether supporting serendipitous discovery can help readers explore reviews more comprehensively, how their exploration behaviors change with access to their exploration patterns, and how suggestions of unexplored reviews might impact their decision-making. 
We formulated the following hypotheses to answer these questions:

\begin{enumerate}
    \item \textbf{H1 - Comprehensive}: Participants who had access to the exploration metrics will cover more reviews.
    
    \item \textbf{H2 - Unbiased}: Participants who had access to the exploration metrics will read a more balanced distribution of reviews. 
    
    \item \textbf{H3 - Confident}: Participants who had access to both the exploration and suggestions will have greater confidence in their decision.
\end{enumerate}

To evaluate our hypotheses, we analyzed the collected quantitative and qualitative data from 100 crowd workers across four different conditions (B, M, S, M\&S). We present the findings from our analysis in this section.\\

\noindent\textbf{H1: Participants who had access to the exploration metrics covered more reviews.} 
We collected the number of reviews across different products that the participants had covered explicitly or implicitly. 
We posited that the participants had explicit knowledge about any review that they visited and implicit knowledge about all reviews that were semantically similar and redundant to the reviews they had explicit knowledge about. 
Fig.~\ref{fig:Serendyze_usage} presents the number of reviews covered on average across all conditions and suggests that the participants who used conditions (M and M\&S) with exploration metrics covered more reviews on average ($203 \pm 77$ and $234 \pm 114$) compared to the conditions (B and S) without exploration metrics ($66 \pm 64$ and $113 \pm 61$). 
The average coverage to the time-spent ratio for conditions B, M, S, and M\&S are $8.64$, $17.28$, $6.72$, and $12.28$, respectively. 
These ratios show that the participants who used conditions M and M\&S had a higher average coverage to time-spent ratio than those who used conditions B and S. 
A higher coverage to time-spent ratio suggests that the participants spent less time covering more reviews.
As such, our results suggest that the participants who had access to exploration metrics covered more reviews efficiently by spending less time to gain more knowledge about the products. 

\begin{table*}
\centering
\scriptsize
\caption{Analysis of reviews covered across conditions. The results of a two-way ANOVA where Conditions and Products are the independent variables and the number of reviews covered is the dependent variable shows a statistically significant difference (p<.05) across conditions but no significant difference across products nor the interaction between conditions and products. Post-hoc Tukey tests indicate a statistically significant pairwise difference in the average number of reviews covered among all pairs of conditions where one condition provides exploration metrics, and the other does not.}
\setlength\tabcolsep{13.5pt}
\ra{1.5}
\begin{tabular}{l c c c c}
	\toprule 
	\textbf{Factors} & \textbf{Degree of Freedom} &\textbf{Mean Sum of Squares} & \textbf{F-value} & \textbf{Pr(>F)}\\
	\midrule
        Conditions & $3$ & $51042$ & $39.21$ & \cellcolor{grayback}$\mathbf{2e^{-16}}$\\
        Products & $2$ & $3177$ & $2.44$ & $0.09$\\
        Conditions : Products & $6$ & $759$ & $0.58$ & $0.75$\\
    \midrule
    \textbf{Condition Pairs} & \textbf{Difference} &\textbf{Lower-bound} & \textbf{Upper-bound} & \textbf{P-adjusted}\\
	\midrule
        Metrics (M) - Baseline (B) & $45.93$ & $30.71$ & $61.16$ & \cellcolor{grayback}$\mathbf{2e^{-11}}$\\
        Suggestions (S) - Baseline (B) & $15.69$ & $0.47$ & $30.92$& \cellcolor{grayback}$\mathbf{.04}$\\
        Metrics and Suggestions (M\&S) - Baseline (B) & $56.16$ & $40.93$ & $71.39$ & \cellcolor{grayback}$\mathbf{5e^{-12}}$\\
        Suggestions (S) - Metrics (M) & $-30.24$ & $-45.47$ & $-15.01$ & \cellcolor{grayback}$\mathbf{2e^{-11}}$\\
        Metrics and Suggestions (M\&S) - Metrics (M) & $10.22$ & $-4.99$ & $25.45$ & $\mathbf{.31}$\\
        Metrics and Suggestions (M\&S) - Suggestions (S) & $40.47$ & $25.24$ & $55.69$ & \cellcolor{grayback}$\mathbf{4e^{-8}}$\\
	\bottomrule
\end{tabular}
\label{tab:h1anova}
\end{table*}

To evaluate \textbf{H1}, since the coverage value for all conditions passed the Shapiro-Wilks test, we performed a two-way ANOVA test with two factors --- conditions and products. 
Table~\ref{tab:h1anova} presents the results of the test. 
The results suggest a statistically significant difference in average reviews covered by participants across conditions.
Furthermore, there are no statistically significant differences among different products and the interactions between the conditions and the products. 
A Tukey posthoc test revealed statistically significant pairwise differences between the conditions that provide exploration metrics (M and M\&S) and the conditions that do not (B and S). 
We posit that the pair Metrics and Suggestions (M\&S) - Metrics (M) are not statistically significant because they both provide exploration metrics. 
Based on the results, we conclude that people with access to exploration metrics covered more reviews, and consider \textbf{H1} to be supported. 
The participants' responses also corroborate these results. Participants in condition M explained how they used the exploration metrics: P31 mentioned, \blockquote{\emph{I used exploration metrics to review enough to make a satisfying estimate of the value of each product. [Coverage] showed me which reviews I hadn't covered yet, so I could read [those reviews] to get a better opinion.}} 
Another participant (P45) said, \blockquote{\emph{I wanted to read a good amount of all kinds of reviews -- positive, neutral, and negative. The Distribution helped me see if I was doing that. [\ldots] I also used Coverage to save time so I didn't read too many redundant reviews.}} 
The participants who used condition M\&S highlighted how having access to exploration metrics helped them read reviews more comprehensively before making decisions. 
P90 said, \blockquote{\emph{The exploration metrics helped me to see what percentage of reviews I had really read to see if I was getting a full picture or not. [\ldots] It helped me to make sure that I know about enough reviews before making a decision.}}
Another participant (P85) mentioned \blockquote{\emph{I used exploration metrics to make sure I was looking at the various types of reviews, including neutral and negative. I wanted to make sure I had read enough of each of these types to make a final decision.}} \\

\begin{figure}
\includegraphics[width=1\textwidth]{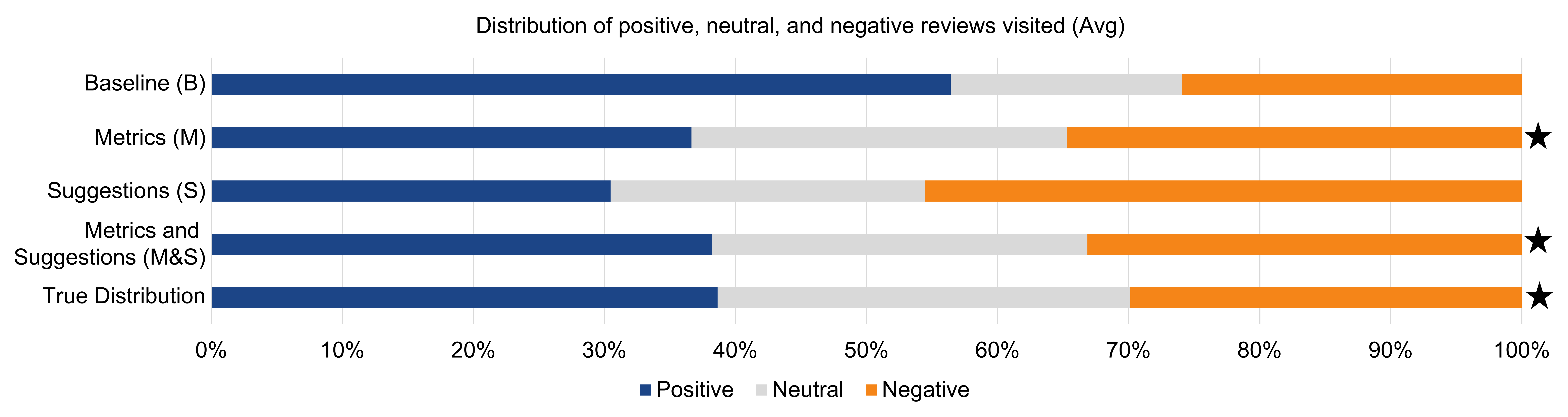}
\caption{This figure depicts the average distribution of positive, negative, and neutral reviews as visited by participants. The figure suggests that participants in conditions M and M\&S, annotated by ($\star$), explored different sentiments in a balanced manner, which is reflective of the true distribution of sentiments present in the dataset. However, in the other two cases, we see an imbalance where participants in condition B visited a relatively larger number of positive reviews, and the participants in condition S visited a relatively larger number of negative reviews.}
\label{fig:Serendyze_dist-perc}
\end{figure}

\noindent\textbf{H2: Participants who used exploration metrics read positive, neutral, and negative reviews in a balanced way.} For each participant, we collected the number of positive, neutral, and negative reviews they visited.
The distribution of the percentage of positive, neutral, and negative reviews visited by participants is presented in Fig.~\ref{fig:Serendyze_dist-perc}.
The figure indicates a notable difference between conditions. The participants who used conditions M and M\&S with exploration metrics visited reviews from all sentiments in a balanced manner, reflecting the true distribution of sentiments present in the dataset used. 
Among 2411 reviews visited by participants in condition M,
on average, each participant visited 35 $\pm$ 24 positive, 27 $\pm$ 15 neutral, and 33 $\pm$ 19 negative reviews. 
The participants in condition M\&S visited 3268 reviews where on average, each participant visited 49 $\pm$ 44 positive, 37 $\pm$ 24 neutral, and 43 $\pm$ 23 negative reviews. 
In contrast, the participants in condition B visited 521 reviews. 
On average, 12 $\pm$ 13 were positive, 4 $\pm$ 8 were neutral, and 5 $\pm$ 7 were negative reviews.
Finally, among 1067 reviews visited by participants of condition S, on average, each participant visited 11 $\pm$ 11 positive, 9 $\pm$ 7 neutral, and 17 $\pm$ 14 negative reviews.
Fig.~\ref{fig:Serendyze_dist-perc} shows that without access to exploration metrics, participants visited more positive (B) or more negative (S) reviews. 
Despite not mentioning the term ``balanced'' during the tutorial, task assignment, and questionnaires, the qualitative responses for participants suggest that the exploration metrics helped participants explore reviews in a more \textit{balanced} way --- as termed by the participants. 
15/25 participants who used condition M and 18/25 participants who used condition M\&S mentioned that they used exploration metrics to balance out how they were reading different kinds of reviews so that they did not lean towards one specific sentiment. 
P28 (condition M) mentioned \blockquote{\emph{I did not want to look at only negative reviews. I used the exploration metrics to make sure I was reading enough reviews of each type.}} 
P49 remarked, \blockquote{\emph{I made sure that I have read a fairly even distribution of all kinds of sentiments, not just positive or negative.}} 
P100, who used condition M\&S mentioned how the exploration metrics enabled them to notice an imbalance in their work and seek out other reviews: 
\blockquote{\emph{I noticed that I had read a lot of positive reviews, so I then read some neutral and negative reviews to balance it out so I got a fuller picture.}}
P77 said, \blockquote{\emph{It was good to be aware that I was viewing a range of review types, and not solely focusing on only positive or only negative reviews.}} 
These findings suggest that our intervention enabled participants to overcome \textit{oversensitivity to consistency}~\cite{heuer1999psychology, wall2017warning} and allowed them to explore all sentiments. 
However, 4/25 participants who used condition M and 3/25 participants who used condition M\&S decided not to use exploration metrics. 
While two of them (P32, P84) mentioned that they \blockquote{\emph{did not feel the need to}}, 4 other participants mentioned (P36, P44, P50, P88) that they, \blockquote{\emph{explored the data on their own, using their own strategy.}} \\

\begin{figure}
\includegraphics[width=1\textwidth]{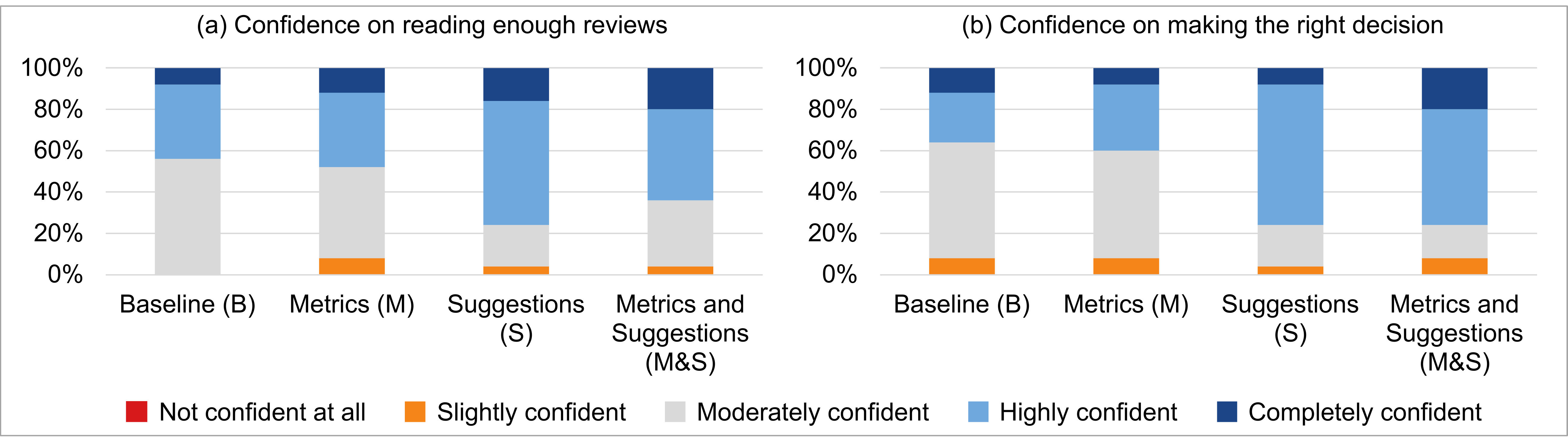}
\caption{This figure shows participants' confidence (a) in reading enough reviews prior to decision-making and (b) in making the correct decision for all Serendyze conditions (B, M, S, M\&S). The figure suggests that in both cases, for conditions S and M\&S, they were more confident in reading enough reviews and on their decisions compared to other conditions.}
\label{fig:Serendyze_confidence}
\end{figure}

\noindent\textbf{H3: Participants who used both exploration metrics and suggestions were confident in their decisions.} 
To understand how exploration metrics and bias mitigating suggestions impact participants' decision-making process, we asked each participant how confident they were about reading enough reviews and making the right decision in the post-study questionnaire (See Figure~\ref{fig:Serendyze_confidence}). 
76\% of the participants (19/25) in condition M\&S were highly or completely confident that they had read enough reviews.   
Furthermore, 64\% of the participants (16/25) in condition M\&S and 76\% of the participants (19/25) in condition S were also highly or completely confident that they had made the right decision. 
However, we did not find similar high confidence among participants who used conditions B or M. 
This result suggests that bias mitigating suggestions might have played a role in invigorating participants' confidence in reading enough reviews and making the right decision. 

\begin{table*}
\scriptsize
\centering
\caption{This table presents the pairwise Mann-Whitney U test results between condition M\&S and the other three conditions (B, M, and S). The results indicate a statistically significant pairwise difference in the confidence among condition M\&S and conditions B and M at an alpha of $.05$. While condition M\&S provides both exploration metrics and suggestions based on the bias mitigation model, conditions B and M do not provide suggestions.}

\setlength\tabcolsep{8pt}
\begin{tabular}{l c c}
	\toprule 
	\textbf{Condition Pairs} & \textbf{z-score} & \textbf{p-value}\\
	\midrule
        Metrics and Suggestions (M\&S) - Baseline (B) & $-2.11$ & \cellcolor{grayback}$\mathbf{.03}$\\
        Metrics and Suggestions (M\&S) - Metrics (M) & $-2.11$ & \cellcolor{grayback}$\mathbf{.03}$\\
        Metrics and Suggestions (M\&S) - Suggestions (S) & $-0.48$ & $.62$\\
	\bottomrule
\end{tabular}
\label{tab:h3anova}
\end{table*}

To evaluate \textbf{H3}, we performed a Kruskal-Wallis rank-sum test, which is a non-parametric test on the distribution of confidence level among participants, since confidence level distribution for all conditions failed the Shapiro-Wilk normality test. 
The results suggest a statistically significant  confidence difference among the participants who used different conditions (B, M, S, and MS). 
For four conditions, the degree of freedom was 3, the critical Chi-Squared value was $10.05$, and the p-value was $p=.02<.05$. 
Hence, we rejected the null hypothesis and followed up by performing a pairwise Mann-Whitney U test with condition M\&S against other conditions (B, M, and S). 
The result of this test is presented in Table~\ref{tab:h3anova}. 
The statistically significant pairs are highlighted in boldface with a gray background. 
The results indicate a statistically significant difference in the confidence of participants on making the right decisions between those who used condition M\&S compared to the participants who used condition B ($p=.03$) or M ($p=.03$). 
However, there is no significant difference among the confidence of participants who used condition M\&S compared to participants who used condition S ($p=.62$). 
This lack of statistical significance further supports the observation that the suggestions based on the bias mitigation model might influence participants' confidence when making decisions. 
Based on these results, we consider \textbf{H3} to be partially supported. 
Although we did not account for it in the study, the pre-study questionnaire suggests that 8/25 participants had either Sony (3 participants) or Sennheiser (3 participants) brand preference. 
However, the post-study questionnaires suggest only one participant from each group decided to select the headphone of their preferred brand, further suggesting that participants could overcome \textit{persistence of impressions based on discredited evidence}~\cite{heuer1999psychology, wall2017warning} and make more confident decisions that do not mirror their preconceived preferences. 

Participants' feedback also suggests that using exploration metrics and suggestions improved their confidence in their decisions. 
P87 (condition M\&S) mentioned, \blockquote{\emph{...[Serendyze] allowed me to process a lot of information quickly. I could search for a specific feature for each headphone product and feel confident about it because of the large amount of positive and negative reviews.}} 
P83 contrasted their experience of using Serendyze with their regular product review patterns, saying, \blockquote{\emph{I appreciated the Distribution a lot. I am very guilty of reading reviews that back up my existing opinion - justifying a purchase rather than really learning about the product... [Exploration metrics] helped me avoid that.}} P93 mentioned how suggestions helped them understand reviewers' perspectives, saying, \blockquote{\emph{It was a quick way for me to see how others felt about the products, and they gave me more information based on the other reviews that I have already read.}} P81 also mentioned how suggestions enabled them to make the right decision by helping them compare between products: \blockquote{\emph{The suggested reviews let me pro and con better as I made my decision.}}\\

\begin{figure}
\includegraphics[width=1\textwidth]{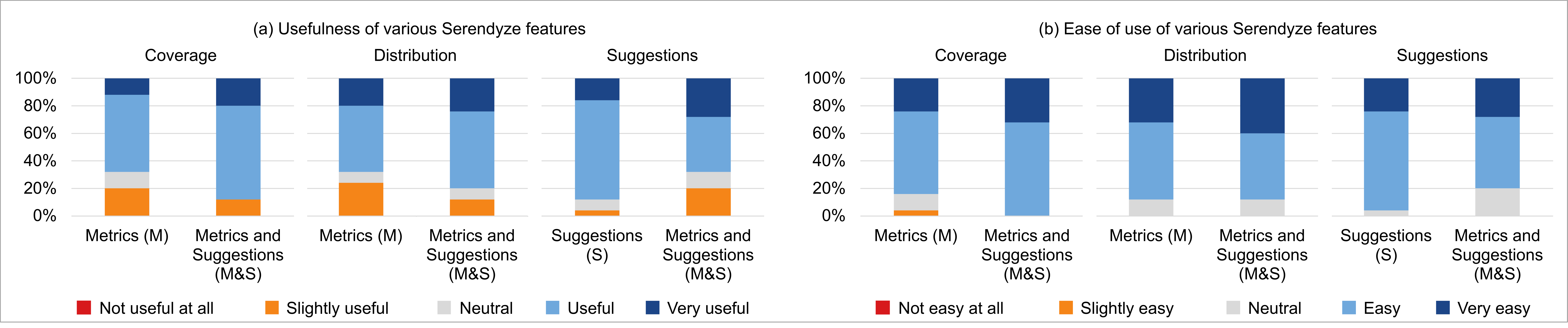}
\caption{This figure depicts how the participants perceived the (a) usefulness and (b) ease of use of Coverage, Distribution, and suggestions based on the conditions where the participants had access to these features. The figure suggests that the usefulness of all features is over 68\%, and ease of use is over 80\% across all relevant conditions.}
\label{fig:Serendyze_usefulness}
\end{figure}

\noindent\textbf{Participants found suggestions to be useful but had mixed feelings about unexpected suggestions.}
During analysis of the post-study questionnaire, we found that the participants found suggested reviews useful for making decisions. Fig.~\ref{fig:Serendyze_usefulness}(a) presents how participants perceived the usefulness of different features provided in Serendyze. 
The figure suggests that 80\% of participants (20/25) who used condition S and 68\% of participants (17/25) who used condition M\&S found the suggestions to be useful or very useful. Responses from the post-study questionnaire suggest suggestions also impacted participants' decision-making (15/25 for condition S and 13/25 for condition M\&S). 

We used qualitative responses collected from participants to understand why and how the suggestions helped participants before decision-making. 
A considerable number of participants in conditions S (13/25) and S\&M (12/25) believed suggestions helped them to gain deeper knowledge from reviews.
For example, P65 mentioned \blockquote{\emph{I decided to choose the headphones that I chose because they had high marks in regards to audio quality based off of the suggested reviews that I was shown.}}
Furthermore, 9/25 participants who used condition S and 6/25 participants who used condition S\&M highlighted that suggestions helped them find unexpected information that they may not have thought about yet. 
P90 (condition M\&S) mentioned \blockquote{\emph{I liked that it was different from what I was reading. They [suggestions] helped me back up my opinion of the product.}} P60 said, \blockquote{\emph{I was suggested reviews I didn't see in the regular one, especially the ones around comfort and styles. I didn't think about those at the beginning.}}
Other participants (5/25 from S and 9/25 from M\&S) mentioned that they found the suggested reviews useful for gaining perspectives on opposite opinions and combat biased exploration. 
P77 mentioned how they leveraged suggestions to get opposite viewpoints, saying, \blockquote{\emph{I would look to the suggestions when I was done reading a particular review and I wanted to read the opposite view. It helped to make sure I didn't get biased towards a product.}} 

\begin{figure}
\includegraphics[width=0.9\textwidth]{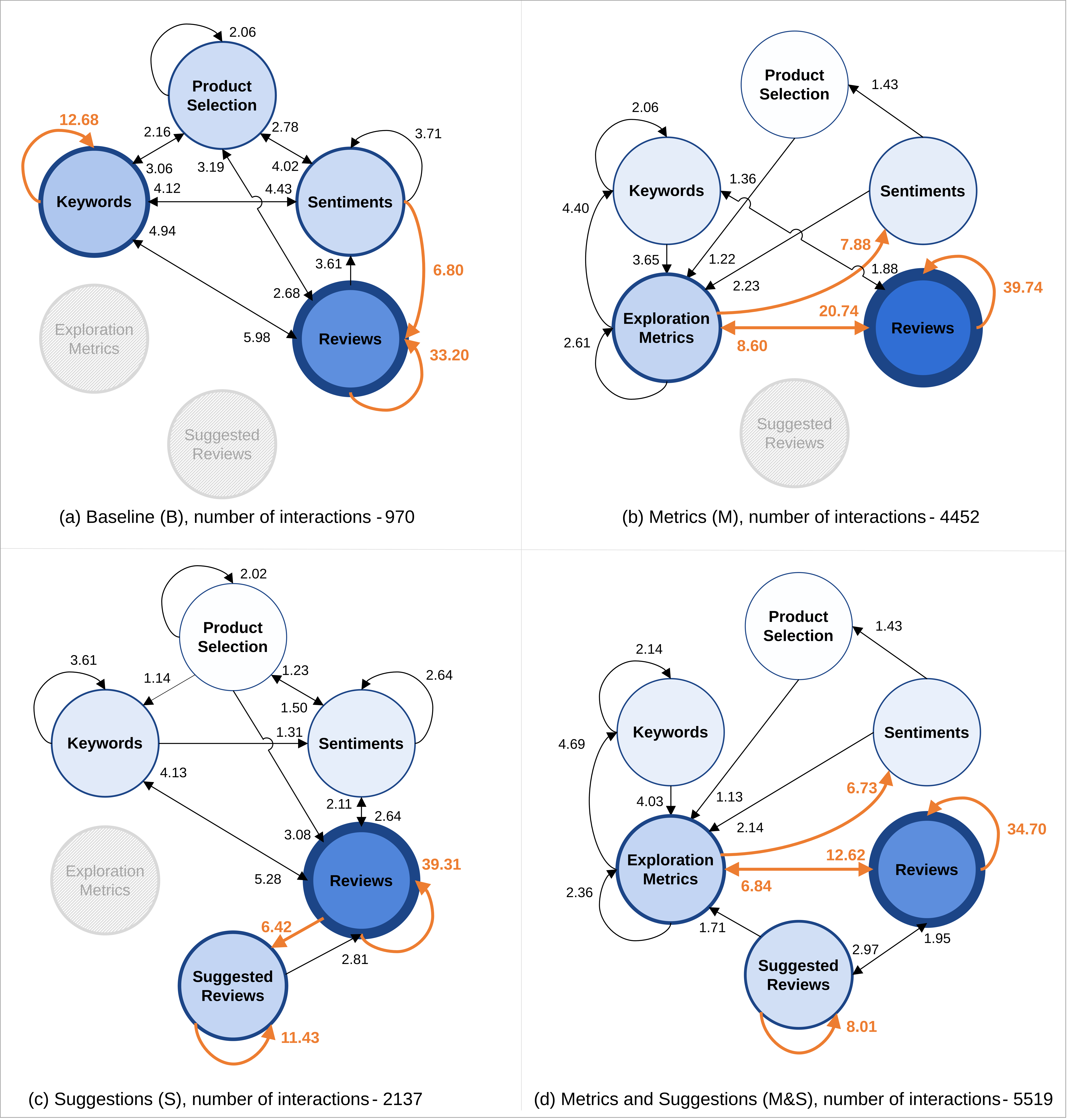}
\caption{Node-link diagrams that show how participants interacted among the six major Serendyze components across all conditions --- product selection, keywords, sentiments, exploration metrics, reviews, and suggestions. Arrows depict a transition from one component to another. For brevity and clarity, the connections between two components are shown only when the number of interactions among them is more than 1\% of all interactions for that condition. The overall use of components is double encoded in the border thickness and background saturation.  The orange lines show connections with a higher frequency of interactions. The figures suggest that participants heavily used exploration metrics and suggested reviews whenever available.}
\label{fig:Serendyze_interaction-track}
\end{figure}

However, not all participants preferred the unexpectedness of the suggestions generated by the bias mitigation model.  
It should be mentioned that during the introduction, tutorial, task assignment, and pre-study questions, we did not mention to participants how the suggestions were generated. 
In the conditions (S and M\&S) where participants had access to suggested reviews, the interface provided reviews as only the suggested reviews \textit{they may like} (see Fig.~\ref{fig:Serendyze_interface}).
Diving deep into the responses of participants who were taken aback by the unexpected suggestions, we found that they often considered the suggestions unhelpful or not detailed enough. 
P76 mentioned, \blockquote{\emph{It [suggestions] gave me unhelpful reviews that made me not to trust the system.}} P61 explained, \blockquote{\emph{suggested reviews were mostly comprised of short reviews [$\ldots$]} that could have been easily generated by bots, so I didn't trust them.}

Some other participants did not find the suggestion useful because they thought that \blockquote{\emph{it did not do a good job}} (P76), or they \blockquote{\emph{could not figure out why they [suggestions] are being suggested}} (P80). These responses suggest that some participants may have had different expectations from the suggested reviews feature when using Serendyze. These expectations could have been an artifact of a priming effect as people often expect recommendations or suggestions to be similar to what they are exploring rather than offering diversity~\cite{kaminskas2016diversity}. 
We further discuss this observation and its ramifications in Section~\ref{sec:discussion}. \\

\noindent\textbf{Participants heavily Used Serendyze interventions to perform text-level analysis of reviews.} 
Figure~\ref{fig:Serendyze_usefulness} suggests that the majority of participants who had access to exploration metrics and suggestions found them to be useful and easy to use. 
However, we wanted to explore deeper and learn how the participants leveraged these features to learn more from the data prior to decision-making. 
We used time-stamped interaction logs to analyze participants' use of Serendyze features to model their exploratory and decision-making strategies (see Fig.~\ref{fig:Serendyze_interface}). 
We should emphasize that while designing the tutorial, task description, and pre-study questions, we paid careful attention to not bias participants towards using any particular feature. The product review task, identical across all conditions, asked participants to explore the reviews and decide which headphones they would refer to someone.

Four node-link graphs in Figure~\ref{fig:Serendyze_interaction-track} show the participants' usage of and transitions between the six primary components of Serendyze --- product selection, keywords, sentiments, exploration metrics, reviews, and suggestions ---for conditions B, M, S, and M\&S). 
The total number of interactions were 970 for the baseline condition (B), 4452 for the condition with exploration metrics only (M), 2137 interactions for the condition with suggestions only (S), and 5519 for condition M\&S, which contains both exploration metrics and suggestions. 
It is worth mentioning that the operational granularity of these interactions is not symmetric. 
For instance, from an interaction perspective, interacting with a keyword might impact a set of reviews but interacting with a review impacts only that particular review. 

Fig.~\ref{fig:Serendyze_interaction-track}(a) suggests that the participants who used condition B often used keywords and sentiments to filter reviews.
The majority of the participants (13/25) preferred the cascaded filters to filter reviews by a keyword and a sentiment in conjunction. One participant (P12) mentioned \blockquote{\emph{I loved the keyword filter to focus on what was important to me and I could then see exactly how many positive and how many negative for that particular keyword. That was amazing!}}
The participants (10/25) also preferred the option to search custom keywords and the highlighting of the searched keyword on the reviews. 
P19 said, \blockquote{\emph{I really liked the way that you could easily search for sentiments. For example, I cared most about reasonable pricing and sound quality. It was easy for search and have those things highlighted in individual reviews.}}

Fig.~\ref{fig:Serendyze_interaction-track}(b) suggests a prominent interaction trend where the participants transitioned between exploration metrics and reviews. 
There are also traces of interactions from exploration metrics to sentiments. 
In essence, the exploration of reviews by participants who used condition M seems to have revolved around exploration metrics as 15/25 participants mentioned they used exploration metrics to balance the types of sentiments they were exploring from the reviews. We found similar trends in condition S (Fig.~\ref{fig:Serendyze_interaction-track}(c), where participants explored both the suggested and regular reviews and went back and forth between them. This observation is also supported by the participants' post-study feedback soliciting how they used the suggested reviews. 
13/25 participants mentioned that they started reading regular reviews; after a while, they started to read suggestions and kept alternating between the two. Finally, in Fig.~\ref{fig:Serendyze_interaction-track}(d), we see that the trends from conditions M and S repeat with the participants having access to all features of Serendyze. These diagrams suggest that the participants heavily used the proposed interventions to perform their tasks whenever available. 

The qualitative responses we collected from the participants also reflect their desire to use exploration metrics and suggestions. 
For instance, P93 mentioned, \blockquote{\emph{Serendyze provides a nice collection of useful information and features in order to compare products. It is something I would use while looking for products online.}} 
P61 highlighted how suggested reviews helped them to keep track of reviews that were important, \blockquote{\emph{The feature that I liked the most was the tracking of the "Suggested reviews that you have visited already." It was really convenient for me to keep track of the reviews that had made an impression on me. It was really easy to add reviews there, and this was important because I've found that it is easy to lose track of specific reviews with a really important detail that was not in other reviews, and this feature is a great solution to prevent losing track of important or personally useful reviews.}} 
P30 mentioned how exploration metrics helped them to learn how much data they have explored, \blockquote{\emph{I used these features [exploration metrics] now and then to get an idea of what information I had already covered. I found them to be very interesting and something I'd like to see on all review pages!}}
The responses for the aesthetics of the Serendyze interface were mixed. 
While some participants (P44, P49) preferred the \blockquote{\emph{color coded interface,}} others (P43, P29) though the interface was \blockquote{\emph{plain and needed more color and designs.}}

\section{Discussion}
\label{sec:discussion}

The findings from the evaluation of Serendyze suggest that exploration metrics helped participants to explore reviews more comprehensively and cover more reviews.  
It also allowed them to balance their review exploration across all sentiments and perspectives present in reviews. 
We also found that the bias mitigation model provided participants with useful suggestions, enabling them to gain deeper knowledge about the products and raising their confidence in making informed decisions. 
Participants also found unexpected suggestions that often impacted their decision by providing evidence. 
However, some participants did not find suggestions useful due to expectation mismatch. 
Overall, the evaluation highlights the usefulness of text-level exploration functionalities to complement summary-level exploration of product reviews to support data-driven decision-making for purchasing online products. In this section, we further unfold the findings from the evaluation, discuss implications of such findings, and speculate how knowledge gained from this work can be propagated to domains beyond product reviews.  

\subsection{The Impact of Preconception and Expectation Mismatch on Accepting Suggestions}
The findings from our evaluation of Serendyze demonstrate that the participants in conditions S and M\&S found suggestions to be useful (see Fig.~\ref{fig:Serendyze_usefulness}(a)) and a catalyst for increasing confidence in making decisions (see Fig.~\ref{fig:Serendyze_confidence}(b)). 
However, some participants (4/25 each for S and M\&S) did not feel that the suggestions were useful as they expected suggestions to be similar to the reviews they visited. 
These participants who were seeking similar reviews from the suggestions were disillusioned and felt disconnected, leading towards their eventual disinterest and discontinuation of reading suggested reviews. 
In contrast, others who embraced the unexpectedness found the suggestions to be serendipitous, compelling, and conducive to gaining insights from the reviews.

In this work, we experimented with a heuristic bias mitigation model to generate suggestions that are most dissimilar to what participants visited --- both in terms of semantic similarity and the sentiments associated with the reviews. 
Such approaches that focus beyond accuracy metrics to measure the utility and performance of generating suggestions are relatively new~\cite{kaminskas2016diversity, shi2012adaptive}. 
Similar to Serendyze, the inner workings of the majority of the available suggestion-generation systems, such as Amazon product recommendations, Facebook and YouTube content recommendations, and Netflix and Spotify entertainment recommendation, are kept hidden from the users.
Such systems often rely on similarity or relevance among user~\cite{shapira2013facebook} or product attributes~\cite{linden2003amazon} based on a users' exploration patterns to suggest new data~\cite{kaminskas2016diversity}. 
These available systems might play a role in priming and shaping user perceptions and heavily impact how and what the users expect from the suggestions generated by an automated system~\cite{kamehkhosh2017user, ekstrand2014user, kapoor2009sequential}.
The participants who actively sought similar reviews might not have perceived that the suggestions were reaffirming their gained knowledge and this expectation mismatch likely resulted in their disconnection from using the dissimilar suggestions. 

To address such issues, systems such as Serendyze could be designed to have the functionality to alternate between providing suggestions that are homogeneous to what the readers have been reading and suggestions that are dissimilar to what they have been reading.
The first approach could help them reaffirm their decisions based on what they read, and the second approach could help them gather serendipitous, broader, and diverse knowledge. 
Such duality might even provide flexibility for users to explore the data as they prefer. 
Furthermore, more clarifications can be added to explain how suggestions are generated. To that end, visual indications of scores and ranks associated with suggestions~\cite{pereira2020rankviz} might improve the readers' perceptions towards suggestions, provide transparency, and combat confusion~\cite{rosenfeld2019explainability}. 

\subsection{Perception of Detailed Reviews and their Impact on Trust in System}
Our evaluation demonstrated how suggestions enabled the participants to gather knowledge from opposing perspectives, access reviews that they had not thought about beforehand, and make confident decisions. 
While suggestions were heavily used by participants (Fig.~\ref{fig:Serendyze_interaction-track}(c) and (d)), some participants were not pleased with suggestions as they did not find them to be helpful.  
For some of these participants, the unhelpful suggestions were sufficient to induce mistrust in the system, and they stopped using suggestions.
Digging deeper, we made two observations: (1) some participants perceived that the suggestions were too brief and did not have enough information to gather deeper knowledge; (2) some considered the lack of details in suggestions to be artifacts of random text generating agents or \emph{``bots''} (P61). 

People often have a complex relationship with how they interact with and use a system, what data they get out of the system, how they interpret such data, and how they establish trust in the system~\cite{chuang2012interpretation}. 
While mistrust of a system can often derive from unexpectedness and uncertainty, the quality of the data provided, and the manner in which they are provided may also have a role to play in how users perceive the data presented to them~\cite{dork2013critical}. 
Serendyze generates suggestions based on the dissimilarities in semantics and the sentiments associated with each review. 
While generating suggestions that intend to support serendipitous discovery and analysis of interesting reviews to diversify knowledge gained from product reviews, the bias mitigation model is not designed to assess the quality of the reviews being suggested. 
Adding quality assessment functionalities to measure the quality of generated reviews could help mitigate the mistrust induced by suggested reviews in systems like Serendyze.

One approach to assess whether the suggestions contain details that might be desirable to the readers could be to modify the bias mitigation model to identify and leverage the latent aspects present in the reviews~\cite{rana2020multi, ding2017neural}. 
For instance, during the scoring of candidate suggestions (see Algorithm~\ref{alg:disinterest}), the model could assess whether certain aspects for a headphone, such as the price, longevity, value, sound quality, or other aspects desirable to the reader, are present in the candidate suggestion. 
Furthermore, the presence and absence of the desired aspects could be provided to readers using visual cues~\cite{isaacs2014footprints, sarvghad2016visualizing} to help readers decide whether they want to read the suggestion. 
Providing cues to missing aspects could also benefit readers by helping them assess whether the suggestions are useful to suit their needs~\cite{sarwar2021utility}.

\subsection{User Agency and Trust in Mixed-Initiative Systems}
For evaluating Serendyze, we asked participants to perform an open-ended task of reading reviews and making a purchase decision. The participants were free to approach the task any way they wanted, and they used Serendyze organically to complete their tasks without provocation to use specific features. 
However, some participants who used condition M\&S (4/25) chose to depend solely on exploration metrics.
They justified their choice by highlighting their preference to not be influenced by machine-generated and algorithmically-curated suggestions and wanted to \blockquote{\emph{analyze the data on their own}} (P98).
They also felt that the visualization of the exploration metrics was the result of their tangible interactions with the system, and the metrics mapped and presented their behavioral patterns without manipulation. 
Such observations open up bigger questions around user agency in mixed-initiative systems such as Serendyze, where users and automated systems work in tandem to achieve a goal~\cite{horvitz1999principles}. 

User agency is a critical concept in mixed-initiative systems, and in human-computer interaction, in general~\cite{yu2017effects}.
It often dictates whether the users will adapt to the functionalities provided by the system~\cite{makonin2016mixed}.
A user is considered to have user agency when they perceive that they are responsible for their interactions with the system and they own the consequences of their actions on the system~\cite{yu2017effects, coyle2012did}. 
Our evaluation suggests that the exploration metrics with the visual cues based on scented widgets~\cite{willett2007scented} enabled participants to feel in control of their exploration process by allowing them to follow their own review exploration strategy. 
This observation aligns with previous works where data visualization has been shown to be effective in conveying information regarding exploratory analysis and open-ended tasks~\cite{hullman2011visualization} to instill a sense of transparency and trust in users~\cite{dork2013critical}.  
In contrast, algorithmic curation of online text --- including and beyond product reviews --- is often associated with a lack of transparency and is conducive to generating mistrust in users due to their closed nature as black-box solutions~\cite{davis2017curation, eslami2017careful}. 

Due to this research area being under-explored, it is challenging to definitively design solutions that balance user agency in mixed-initiative systems. 
However, the debate remains on how to address the volatility of user agency in mixed-initiative systems with inevitable black-box components and algorithmically curated system responses.
One might argue that the system should enable users to have agency and have the capability to support users' choices of rejecting features that they do not feel comfortable adapting to. 
Others might advocate providing additional features and guidance to make automated systems more transparent~\cite{ceneda2021show}. 
These questions and viewpoints demand the attention of researchers from multiple disciplines including human-computer interaction, machine learning, and visual analytics. We extend the call to future researchers to investigate these questions and devise solutions to how the dichotomy between user agency and trust in mixed-initiative systems can be balanced.  

\subsection{Approaches such as Serendyze can Facilitate Deeper Fine-Grained Knowledge Acquisition}
From our evaluation, we found that the participants were keen to use review-level analysis features available to them to read and analyze reviews in detail. Apart from the exploration metrics and suggestions, three features that were prominently mentioned by participants across all conditions are: (1) the ability to search for any keyword they wanted and to see them highlighted in the filtered reviews, (2) the ability to filter reviews by different sentiments (positive, neutral, and negative), and (3) the provenance tracking where they could mark the reviews they read. 
The free-form keyword searches and highlighting provided users with the freedom to explore the reviews by focusing on what is important for them. 
The sentiments gave readers a nuanced sense of reviewers' disposition towards a product, which is different from visually presenting star ratings, as star ratings may not best reflect the affinity represented in reviews~\cite{schreck2019online}.
Finally, the provenance tracking enabled them to track their review exploration without the need for mental notes, reducing cognitive effort for decision-making.
Participants across different conditions expressed their desire to see functionalities provided in Serendyze on \emph{``Amazon''} (P12, P49) or similar \emph{``online sites''} (P18, P97). 

Two overarching insights can be extracted from this observation. 
First, our participants' interactions with reviews suggest a lack of available functionalities and options to analyze reviews. 
Major online commerce websites (Amazon, Etsy, eBay, etc.) host numerous products with thousands of reviews per product, but often do not provide powerful features to analyze reviews directly as texts. 
While there are filters such as price range, warranty, color, etc., these are product-level filters enabling analysis among products based on metadata attributes.
They are often not connected with reviews for the product, and the readers seeking to purchase a product based on others' reviews have to painstakingly read through the reviews, often make mental notes, and make decisions based on incomplete knowledge~\cite{qualtrics}.

While Amazon and eBay provide keywords extracted from reviews, the exploration capability they provide is often limited for readers who might have the desire to explore reviews more comprehensively. This leads us to the second insight: the desire for fine-grained analysis at the review level. Participants' appreciation towards these seemingly rudimentary features suggests the utility of review-level analytics where the analysis can be performed on the review contents and highlights the usefulness of integrating such features on available platforms. 

However, such options lead us to questions around identifying the appropriate granularity~\cite{Shneiderman96theeyes} of information to present to readers for exploring and analyzing reviews. Some of these questions involve how to enable readers to analyze review content more efficiently while negating redundancy and how to combine visualization and computational approaches to disseminate information at multiple levels of granularity. 
In the future, researchers from HCI, visualization, natural language processing, and information retrieval could collectively explore paradigms of information seeking when review-level analytics is integrated with summary-level overviews. These paradigms could also explore domains beyond product reviews where text analysis can support decision-making. 

\subsection{Application of Serendyze in Other Domains beyond Product Reviews}
Our study revealed an opportunity for review-level analysis of product reviews to help readers learn more from the data prior to making data-driven purchase decisions. 
This approach could be expanded in domains where comprehensive exploration and text-level analysis of text data could be important to support decision-making. 
One such domain is civics, where decision-makers depend on large-scale public input to gain an understanding of public perception before making critical policy decisions~\cite{jasim2021communitypulse, mahyar2019civic}. 
They often use analytics tools that enable an analysis of public-generated data --- predominantly text data as comments, ideas, and opinions --- to measure the temperature of public perception~\cite{mahyar2019civic}.
While tools designed for analyzing redundant and often ambiguous public input help decision-makers get high-level overviews of public opinions, marginalized and unpopular opinions are often neglected due to the scale of public input and lack of analysis tools to identify such opinions~\cite{mahyar2020designing}, especially at text-level~\cite{jasim2021communitypulse}. 
Since these decisions directly impact peoples' lives, effective analysis to ensure the perspectives of all citizens are addressed is critical in this domain~\cite{mahyar2019civic}. 
Interventions such as the exploration metrics can help decision-makers to identify and extract insights from redundant information and track whether their public input exploration is skewed towards certain agendas, topics, or sentiments. 
Furthermore, the bias mitigation model can suggest opinions and feedback that might have remained hidden under more popular opinions. 
As such, these interventions could provide decision-makers in the civic domain an alternative approach to not only gain a holistic understanding of public input but also enhance their accountability and transparency~\cite{jasim2021communityclick}, when making policy decisions. 

Another domain where text-level analysis systems such as Serendyze can be expanded is social media content analysis. 
While there exists a plethora of tools and techniques to analyze social media texts~\cite{marcus2011twitinfo, hu2016visualizing}, the issues regarding aggregation and summarization of opinions may also manifest in this domain~\cite{wu2016survey}. 
Such issues are especially pertinent due to the concerns around algorithmic filtration and curation of social media content based on users' digital footprints~\cite{bhargava2019gobo, papakyriakopoulos2020political}.
These curating algorithms often decide what social media content the readers should be exposed to~\cite{lambrecht2019algorithmic, bozdag2013bias}, which might result in inadvertently creating filter bubbles~\cite{pariser2011filter}.
For many people who use social media as a source of news and current affairs, such curation and presentation of catered data might promote homophily~\cite{bisgin2012study} and render the readers oblivious to the bigger picture of current affairs in virtual social spaces~\cite{messing2014selective, flaxman2016filter}. 
Text-level analytics systems such as Serendyze can help to combat the formation of echo chambers via serendipitous suggestions of social media content that are dissimilar from the posts that a reader usually explores and are exposed to in social media.
For instance, if a reader is mostly exploring content  from sources aligned with liberal ideas, they could be suggested content from sources that are inclined towards conservative thoughts. 
While social media users will maintain the agency to decide which ideas they align with and own their actions, such text-level intervention can enable them to be introduced to opposing ideas that might help them reach a better understanding of arguments from all sides prior to establishing social alignments. 

\section{Limitations and Future Work}

\textbf{Limitations.} 
One of the limitations of Serendyze is the latency associated with performing the pairwise comparison of visited and unvisited reviews to measure the similarity scores and generate suggestions. 
While the system performed well in a local system and during the pilot study, during the study with crowd workers, with up to 72 participants working simultaneously, the freely available server~\cite{pythonanywhere} used to perform the calculations was overwhelmed with traffic. As a result, some of the participants (6/100) felt that the system worked slower than they expected. 
We emphasize that the latency is an outcome of logistical challenges and could have been mitigated with a more powerful back-end server or batch-wise distribution of tasks among crowd workers. 
In the future, we will optimize Serendyze to perform more efficiently in low-resource environments. 

The other limitation involves the interface and the associated complexity. 
Some participants, especially the ones who used condition M\&S (4/25), found some components of Serendyze to be confusing and to contribute interface clutter.
To mitigate this issue, the Serendyze interface could be improved by enabling participants to hide not just the suggestions but any component that they might not want to see. 
Although the inner workings of generating suggestions were not explained to participants due to study purposes, in the future, the participants can be informed by adding an explanation to remove confusion and increase transparency. 

Serendyze is designed as a customizable and modular web application. For this study, instead of probabilistic machine learning approaches, we used deterministic approaches to analyze reviews that included using keywords extractions based on co-occurrence and using star-rating as the foundation for sentiments.
We emphasize that Serendyze can be outfitted with advanced computational methods to generalize it for tasks and domains where probabilistic classifications are acceptable and desired for scalability.
However, in this study, we focused more on the interaction design and less on the computational approaches.
As such, we adopted deterministic approaches to identifying keywords, sentiments, and similarities among reviews. 

In our study, we only recruited participants who resided in North America via Amazon Mechanical Turk and did not account for participants' demographic information.  We focused on people's purchase practices and experiences irrespective of their backgrounds. Furthermore, the study was limited to a single session which could have impacted some participants in accelerating the decision-making process. In the future, we plan to deploy Serendyze as a longitudinal study to track participants' purchase behaviors over a month across multiple sessions on multiple online products and among participants from diverse demographics. 
Such experiments will enable us to further study the long-term impact of exploration metrics and bias mitigating suggestions on people's review exploration, holistic understanding, and data-driven decision-making based on their purchase habits and experiences that may vary across different regions. 

\textbf{Future Work.}
There are several avenues to explore in the future to improve Serendyze. 
We will study the utility of Serendyze in real-world scenarios by deploying it as a companion web application or browser extension that can enable readers to utilize Serendyze features to explore reviews on online commerce sites. 
In these real-world deployments, Serendyze will be outfitted with product reviews that mirror the real distribution of facets across product reviews.
Before deploying Serendyze in a real-world setting, we will augment it with several functionalities based on this study and the knowledge we gained from the participant responses.
For instance, we will add clarifying information to explain all components and optimize Serendyze to improve the scalability. 
One way to improve the scalability is to use non-tabular databases and pre-calculations to accelerate the query process to measure the dissimilarity scores.  
We will also integrate and enable the readers to hot-switch between different suggestion-generating models to account for their exploration preferences during review exploration. 
Serenedyze's modular and customizable design (see Section~\ref{sec:system}) will allow us to experiment with various text analytics methods to enable exploration of various facets present in the data, including subjectivity~\cite{chaturvedi2018bayesian}, stance~\cite{kuccuk2020stance}, and latent aspects~\cite{rana2020multi, ding2017neural}. 

Some argue that product review distributions in online commerce websites are often inherently biased based on self-selection biases such as purchasing bias and under-reporting bias~\cite{hu2009overcoming}. 
Such biases often result in the review distribution being bi-modal or non-normal, leaning more towards positive or negative reviews~\cite{hu2009overcoming, dave2003mining}. 
While we did not engage with such possibilities in this study, in the future, one avenue to explore is to study the effect of presenting suggestions that negate word of mouth on decision-making~\cite{hu2006can}. 
The modular and customizable design of Serendyze will enable us to replace the bias mitigation model with other statistical models appropriate for such studies. 
We also plan to study people's exploration patterns if they were limited to reading a fixed number of reviews, a fixed amount of time~\cite{wang2021revamp}, or a fixed organization of suggestions. 

In the future, Serendyze could also be outfitted with features to disseminate and allow exploration and analysis of various product and review attributes, including product specifications, pictures, price, warranty information, peer rating, etc. Peer rating could also be used to weigh the suggestions to provide recommendations based on how others valued a product. In addition, Serendyze could be improved by adding features to compare between two or more products in juxtaposition. 
Further improvements can be made by adding note-taking functionalities for the readers to further reduce mental load prior to decision-making. 
The Serendyze interface could also be updated with improved aesthetics and accessibility features to make it more presentable. 

Another avenue to explore in the future is broadening the investigation and assessing the applicability of systems like Serendyze in other domains. 
For instance, in the digital civics domain, exploration metrics and bias mitigating suggestions could help decision-makers identify marginalized or unpopular perspectives among often redundant public-generated data. Furthermore, Serendyze could be used to analyze social media posts of contentious or divergent topics to help combat echo chambers~\cite{cinelli2021echo}.  
In the future, we will collaborate with government and non-government organizations (NGO) --- who collect, analyze, and make decisions based on public-generated data --- to deploy and study how Serendyze could provide them with an alternative to their existing data analysis process by helping them gain deeper insights and a holistic understanding of public-generated texts. 

The exploration metrics and bias mitigating suggestions in Serendyze could also be expanded beyond reviews and text, in general, to other media types, including photos or videos. For instance, in applications such as Yelp, the bias exploration metrics and bias mitigating suggestions might help viewers identify distinct popular dishes, attractions, or places of interest among often redundant photos posted by people who have already experienced these items. 

\section{Conclusion}

In this study, we investigated interventions that are intended to support serendipitous discovery and analysis of product reviews to help readers to explore reviews more comprehensively in a balanced way, prior to making purchase decisions. 
First, we proposed three exploration metrics --- Visit, Coverage, and Distribution. 
These exploration metrics were designed to help readers to keep track of what reviews they have explicitly read, which reviews they have implicit knowledge about, and how they have been exploring different facets of reviews such as sentiments compared to the true distributions of these facets present in the data. 
Second, we proposed a bias mitigation model that generated suggestions based on what the readers had been exploring by identifying and suggesting reviews that were semantically and sentiment-wise dissimilar to the reviews the readers had read already.
This model was designed to generate suggestions that could help readers mitigate biased exploration, guide readers to gain a more comprehensive understanding of the reviews, which was reflective of the true distributions of the semantic and sentiment diversity in the reviews, and enhance their knowledge discovery.
We integrated these interventions with a text analytics system, Serendyze. 
Our evaluation with 100 crowd workers suggests that the exploration metrics could enable readers to cover more reviews in a balanced way. 
We also found that the suggestions generated by the bias mitigation model could be influential in enabling readers to make confident decisions. 
While we do not claim that serendipitous discovery and analysis is the only way to approach purchase decision-making based on online products, the findings from our study suggest that readers seeking to gain a more comprehensive understanding of the underlying reviews might be benefited if they have access to such alternative interventions. 
We discuss the impact of readers' perceptions on accepting suggestions from a system and how user agency in mixed-initiative systems might play a significant role in how users trust interventions that generate guidance for them on what they can or should do while using such a system. 
We also discuss how systems like Serendyze might be useful when expanded to other domains beyond product reviews to support the deeper exploration of text data prior to making data-driven decisions.     


\bibliographystyle{ACM-Reference-Format}
\bibliography{sample-base}

\end{document}